\documentclass[journal]{IEEEtran}

\usepackage{graphicx}
\DeclareGraphicsExtensions{.eps}
\graphicspath{{FigFile/}}

\usepackage{cite}
\usepackage[tight,footnotesize]{subfigure}

\usepackage{subfigure}
\usepackage{color}

\usepackage[cmex10]{amsmath}

\usepackage{epstopdf}
\newcommand{\PhysVec}[1]{\boldsymbol{#1}}

\begin{document}
\title{Design of a Planar Eleven Antenna for Optimal MIMO Performance as a Wideband\\ Micro Base-station Antenna}
\author{Aidin Razavi,
		Wenjie Yu,
		Jian Yang, ~\IEEEmembership{Senior Member,~IEEE,}
		and \\ Andr\'{e}s Alay\'{o}n Glazunov,~\IEEEmembership{Senior Member,~IEEE}
\thanks{	
	A.~Razavi was with the Signals and Systems Department, Chalmers University of Technology at the time this work was conducted (e-mail: aidin.razavi@tgeik.com). He is now with Ericsson AB, Sweden. W. Yu, J. Yang are with the Electrical Engineering Department, Chalmers University of Technology, SE-41296 Gothenburg, Sweden (e-mail: wenjiey@student.chalmers.se; jian.yang@chalmers.se). A. A. Glazunov A.~A.~Glazunov is with the Department of Electrical Engineering, University of Twente, P.O. Box 217, 7500 AE Enschede, The Netherlands and he is also affiliated with the Department of Electrical Engineering, Chalmers University of Technology (e-mail: a.alayonglazunov@utwente.nl; andres.glazunov@chalmers.se).}
}

\markboth{Submitted to journal}%
{Shell \MakeLowercase{\textit{et al.}}: Bare Demo of IEEEtran.cls for Journals}

\maketitle

\begin{abstract}
A new low-profile planar Eleven antenna is designed for optimal MIMO performance as a wideband MIMO antenna for micro base-stations in future wireless communication systems. The design objective has been to optimize both the reflection coefficient at the input port of the antenna and the 1-bitstream and 2-bitstream MIMO efficiency of the antenna at the same time, in both the Rich Isotropic MultiPath (RIMP) and Random Line-of-Sight (Random-LOS) environments. The planar Eleven antenna can be operated in 2-, 4-, and 8-port modes with slight modifications. The optimization is performed using genetic algorithms. The effects of polarization deficiencies and antenna total embedded efficiency on the MIMO performance of the antenna are further studied. A prototype of the antenna has been fabricated and the design has been verified by measurements against the simulations.
\end{abstract}


%
\IEEEpeerreviewmaketitle

\section{Introduction}
\IEEEPARstart{M}{ultiple}-input multiple-output (MIMO) dual-polarized wideband antennas are required in future wireless communication systems, such as in the 5G communication system. The 5G systems will rely on many small cells and micro base-stations~\cite{chen2014requirements}. This will lead to a need for low-profile multi-port antennas with MIMO capability and simple manufacturing process.

The Eleven antenna is a dual-polarized ultra-wideband (UWB) antenna with a decade bandwidth. It has been used as a feed for reflector antennas \cite{olsson2006eleven} and demonstrated good performance in radio telescope applications \cite{yang2009design, yang2011cryogenic}. In addition to the applications for reflector antennas, the multi-port Eleven antenna has been studied for use in, e.g., monopulse tracking systems \cite{yin2007monopulse} and UWB communication systems as a MIMO antenna \cite{yang2010measurements, chen2011comparison}. All these characteristics, make Eleven antenna a suitable choice for base-station antenna.

When dealing with wireless communication systems, it is much desired to perform system level measurements such as throughput and Probability of Detection (PoD), instead of the static antenna characteristics such as radiation pattern, directivity and gain. In these cases, channels are emulated and links are established in an Over-The-Air (OTA) setup, so the statistical system level performance of the wireless system can be evaluated.

The systematic characterization approach is proposed in \cite{kildal2013new} for OTA measurement evaluation of wireless devices. In this approach two extreme reference environments (namely edge environments) are studied. The first edge environment is the Rich Isotropic MultiPath (RIMP) environment where multiple propagation paths are present between the two ends of the wireless link and the channel undergoes Rayleigh fading.  At the receiving side, multipath environment can be emulated by several incoming waves with uncorrelated amplitudes, phases, polarizations and angles of arrival (AoA)  \cite{kildal2015foundations}. The rich isotropic multipath environment is the hypothetical extreme multipath environment defined as a reference, for the convenience of measurement. \emph{Isotropic} refers to uniform distribution of AoA of the incoming waves within $4 \pi$ solid angle, while the term \emph{Rich} means the number of incoming waves is large, typically more than 100 \cite{kildal2013new}. When the intended coverage of the antenna is limited (such as wall or ceiling mounted antennas with half-sphere coverage), Rich MultiPath (RMP) is a more accurate term to use. We herein use RIMP as a general term covering both isotropic and coverage-limited cases. The RIMP environment is usually emulated in a reverberation chamber (RC) which is fitted with reflectors and mode stirrers to generate the rich environment.

The second edge environment is the Line-Of-Sight environment (LOS), where reflections and diffraction in the environment are small. On the other hand, and the direct path between the transmitter and the receiver is unobstructed and there is one dominant path between the two ends of the link. Anechoic chambers (AC), with absorbers fitted on the walls, are traditionally used for emulation  of the LOS environment. Anechoic chambers are mainly used for antennas with directive beams, which are intended for fixed installations. However, in mobile communications, the situation is not static due to the randomness in the orientation in which the wireless terminal is held by the users. To distinguish this situation from the traditional LOS where the antennas on the two sides are fixed, we call this environment Random-LOS. When the distance between the base station and the user is short (the case for micro base-station) or when the operating frequency is high at millimeter waves, the Random-LOS scenario will be more relevant than the RIMP and fixed LOS scenarios.

The real-life propagation environment is a combination of both RIMP and Random-LOS. The edge environments and the real-life scenario are related through a hypothesis stating: \textit{If a wireless device works well in both RIMP and Random-LOS, it will also work well in real-life environment}~\cite{kildal2013new}.

\begin{figure}[t]
	\begin{center}
		\includegraphics[width=0.3\textwidth, height=0.3\textwidth]{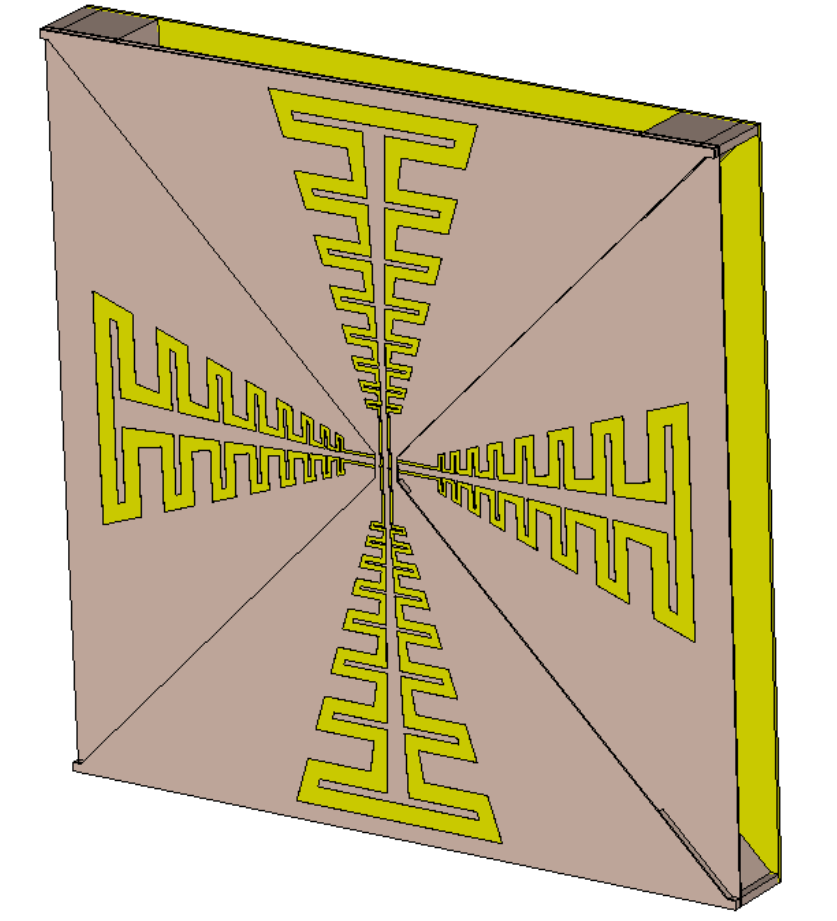}
		\caption{CST model of the present planar MIMO Eleven antenna with a size of $337\times337\times37$ mm$^3$.}
		\label{fig:CSTmodel}
	\end{center}
\end{figure}

In the current paper, we propose a new low-profile planar MIMO Eleven antenna as shown in Fig.~\ref{fig:CSTmodel}, including two branches located in two separate planes corresponding to two different polarizations orthogonal to each other, for the applications in future wireless communication systems. We characterize the performance of the antenna in both RIMP and Random-LOS environments.
However, when it comes to small cell sizes and low powers, multipath fading decreases and Random-LOS will play a larger role. Hence, our focus is on the Random-LOS scenario. The planar Eleven MIMO antenna has a simple geometry and therefore a low manufacturing cost. The design criteria is to optimize both the reflection coefficient and the 1-bitstream and 2-bitstream MIMO efficiency.

In order to analyze the system performance in the edge environments, we use ViRM-lab, a computer program investigating performance of wireless terminals in multipath and LOS with arbitrary incident waves \cite{carlberg2010virmlab}.

A prototype of the antenna has been fabricated and the design has been verified by measurements against the simulations. Simulated and measured results are presented in the paper.

\section{Theory and Figure of Merit}

\subsection{Digital Threshold Receiver Model}
The Probability of Detection (PoD) is the probability that a bitstream is received at the receiver with no errors. It can be described as the normalized throughput of the system. In this paper we employ the \emph{Ideal Digital Threshold Receiver (IDTR) Model}~\cite{kildal2011threshold} in order to obtain the PoD from the probability distribution of the received power. This model was originally introduced to model the throughput of digital communication systems in the RIMP environment. However, it can easily be extended to Random-LOS environment, as well.

The IDTR model which relates probability distribution of the received signal power to the PoD, is based on the simple fact that in modern digital communication systems, the bit error rate in a stationary additive white Gaussian Noise (AWGN) channel, changes abruptly from 100\% to 0\% at a certain threshold signal-to-noise ratio (SNR), due to the use of advanced error correction schemes. The threshold level is determined by the receiver and the performance of the wireless system.

According to the IDTR model, the PoD is determined by~\cite{kildal2011threshold}:
\begin{equation}
\mathsf{PoD}(P/P_\text{th})=\frac{\mathsf{TPUT}(P/P_\text{th})}{\mathsf{TPUT}_\text{max}}=1-\mathsf{CDF}(P_\text{th}/P),
\label{eq:ThReceiver}
\end{equation}
where $\mathsf{PoD}$ is the probability of detection function, $\mathsf{TPUT}$ is the average throughput, $\mathsf{TPUT_\text{max}}$ is the maximum possible throughput (depending on the system specifications), $\mathsf{CDF}$ is the Cumulative Distribuion Function (CDF) of the received fading power ($P_\text{rec}$), $P_\text{th}$ is the receiver's threshold level and $P$ is a reference value which is proportional to the transmitted power and defined according to the environment.

\begin{figure}[!t]
	\begin{tabular}{cc}
		\includegraphics[width=0.45\columnwidth]{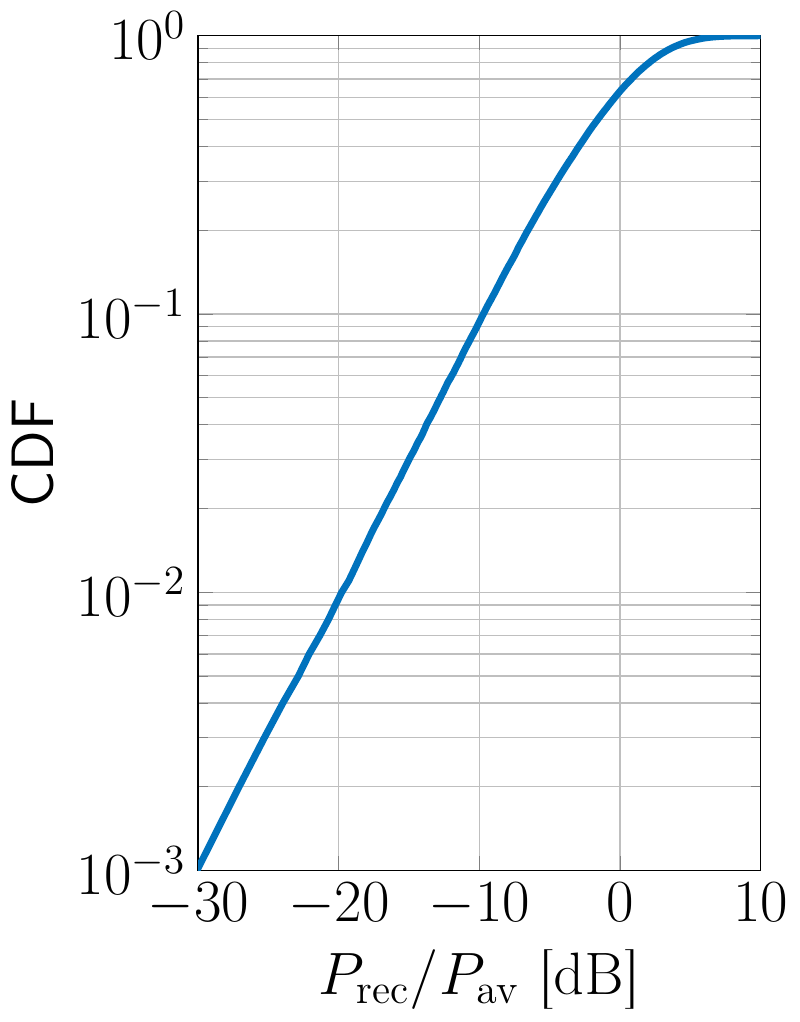}&
		\includegraphics[width=0.44\columnwidth]{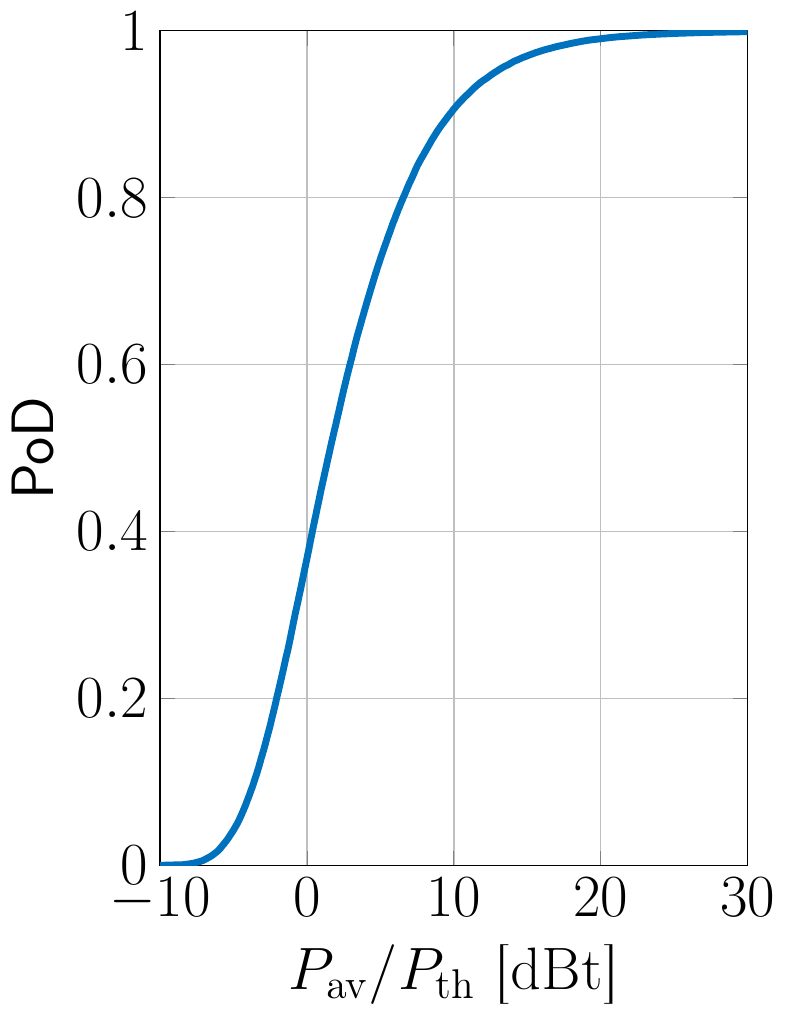}\\
		(a) & (b)
	\end{tabular}
	\caption{Illustration of ideal threshold receiver model for i.i.d. multipath case (a) the CDF, and (b) the PoD.}
	\label{fig:ThReceiver}
\end{figure}

To illustrate the IDTR model with an example, let's assume the example of an isotropic antenna in rich multipath environment (i.i.d. case). The CDF of the received power, which is of a Reyleigh distribution, is plotted in Fig.~\ref{fig:ThReceiver}(a), where the reference power $P$ is chosen as the average received power ($P=P_\text{av}$). This CDF plot shows that, e.g., in 9\% of the states, the received power is at least 10 dB below the reference level $P_\text{av}$. This means that for the remaining 91\% of the states, the received power is not more than 10 dB below the reference level. This means that if the threshold level $P_\text{th}$ is 10 dB below the the reference level, there is 91\% probability that the received power is above the threshold level which means no bit error, so the PoD is 91\% in this case. The PoD of the i.i.d. case with the above-mentioned assumptions is plotted in Fig.~\ref{fig:ThReceiver}(b), which shows that in order to maintain an error-free link for 91\% of the time, the average received power needs to be at least 10 dB above the threshold level, i.e., 10 dBt.

\subsection{MIMO efficiency}
The power required to maintain 95\% PoD level is often used as a metric for the performance of digital communication systems \cite{HussainKildalGlazunov2015, chen2014throughput}. This metric represents the power that is required at the transmitter side so that the receiver can detect 95\% of the data packets, for a fixed coding and modulation scheme. MIMO efficiency is defined by the degradation of $P/P_\text{th}$ at 95\% PoD compared to an ideal receiver antenna. The reference antenna is chosen depending on the intended coverage of the antenna and the considered environment, i.e., RIMP or Random-LOS. MIMO efficiency can be expressed as:
\begin{equation}
\eta_\text{MIMO} = \frac{\mathsf{PoD}_{0}^{-1}\left(0.95\right)}{\mathsf{PoD}^{-1}\left(0.95\right)},
\label{eq:MIMO_efficiency}
\end{equation}
where $\mathsf{PoD}^{-1}$ is the functional inverse of $\mathsf{PoD}$, and $\mathsf{PoD}_{0}$ represents the PoD of the reference antenna. The reference power level $P$ in \eqref{eq:ThReceiver} should be the same for both the reference antenna and the antenna under study, but its actual value does not affect the MIMO efficiency.

We will study the performance of the planar Eleven antenna in both 1-bitstream and 2-bitstream scenarios, and evaluate the MIMO efficiency in both cases. As implied by the names, in the 1-bitstream system, one data stream is transmitted between the two ends of the link, while in the 2-bitstream system two independent data streams are carried in the link. Since most wireless terminals are limited to two antenna ports, we limit this work up to the 2-bitstream case. However at the base station side, where the planar Eleven antenna is located, 2, 4, and 8 antenna ports can be available depending on the configuration. Maximal-Ratio Combining (MRC) and Zero-Forcing (ZF) algorithms are used to combine the multiple antenna ports for 1-bitstream and 2-bitstream cases, respectively.

For a 1-bitstream system, the efficiency corresponds in reality to a SIMO efficiency. However, in the current paper we use MIMO efficiency as a general term covering both SIMO and MIMO scenarios.

\subsection{Reference Antenna}
As mentioned earlier, the choice of the reference antenna depends on the environment. For Random-LOS, the reference is chosen as an isotropic lossless antenna, which is polarization-matched to any random polarization of the incoming wave. This hypothetical reference is an antenna which provides constant output power, regardless of the AoA and polarization of the incoming wave. Similarly in the RIMP case, the reference is chosen as an isotropic lossless antenna in a rich multipath environment. Therefore, the output power at the port of this reference antenna, follows a Rayleigh distribution.

If the intended coverage of the antenna is not the whole sphere, instead of an isotropic antenna, the reference antenna is chosen such that its radiation pattern is uniform in the whole intended coverage and is zero out of the intended coverage. In the case of limited coverage, the amplitude of the reference antenna is adjusted proportional to the solid angle of the intended coverage, so that the total radiated power is the same as the isotropic antenna. The intended coverage area of the current planar Eleven antenna is $120^\circ$ in both azimuth an elevation planes. This coverage area is the same for both RIMP and Random-LOS cases. The coverage area can be described in spherical coordinate system as:
\begin{equation}
\pi/6\leq\theta\leq 5\pi/6,\quad -\pi/3\leq\phi\leq \pi/3.
\label{eq:coverage}
\end{equation}


\section{Modeling and Optimization}
\subsection{Layout}
%

\begin{figure}[t]
	\centering
	\includegraphics[width=0.8\columnwidth]{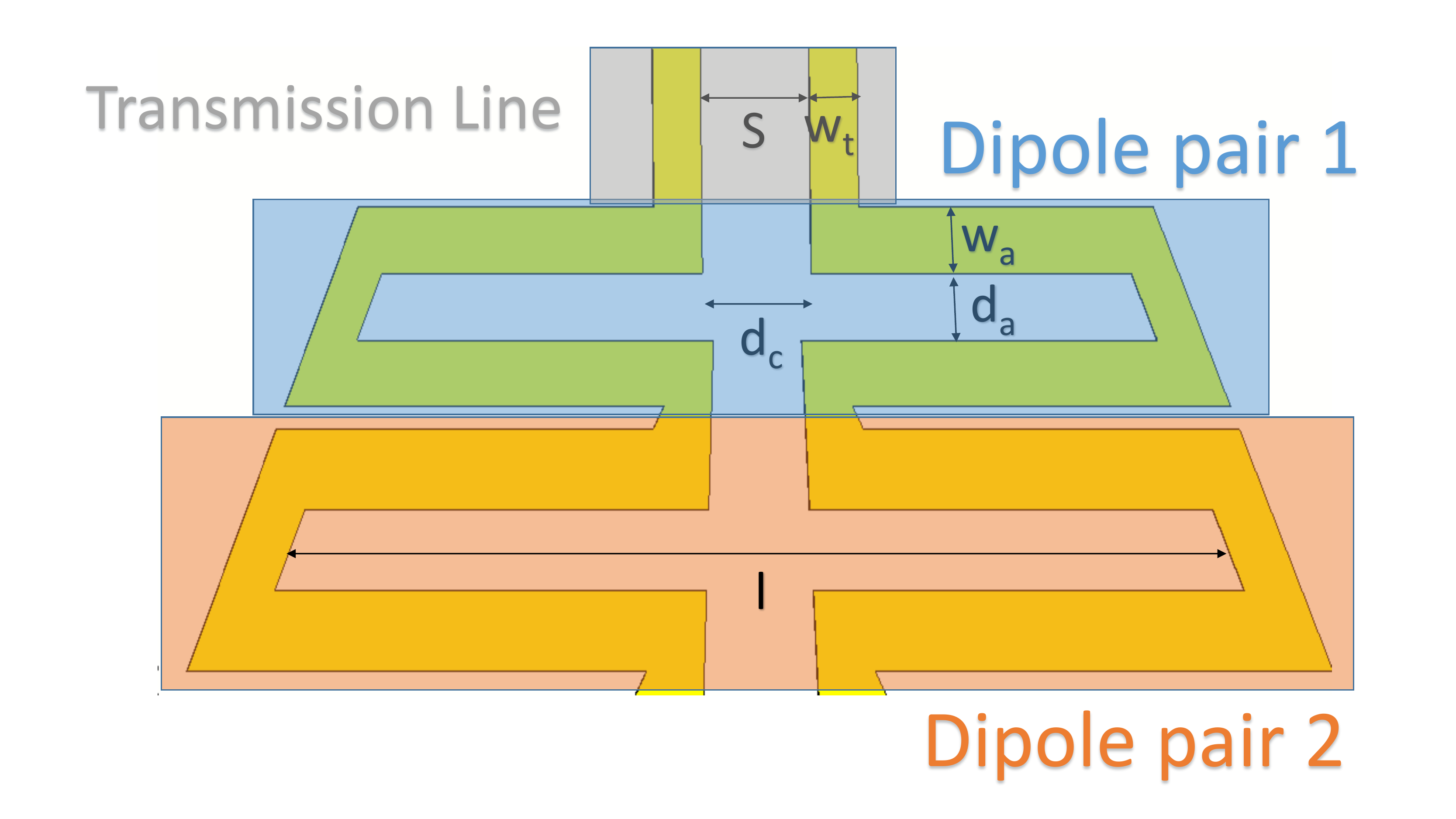}
	\caption{Illustration of design parameters.}
	\label{fig:parameters}
\end{figure}

In order to lower the manufacture cost, we use the flat configuration of the Eleven antenna as shown in Fig.~\ref{fig:CSTmodel}. In this work, we design the prototype for a two-port dual-polarized MIMO antenna with a simple feeding structure at the center. Therefore, the antenna is composed of two orthogonal radiation panels, one at the upper layer and the other at lower layer with a separation of $3$~mm. In fact, if four-port or eight-port antennas should be used, the two-layer structure can become one layer with even much lower manufacture cost and simpler feeding structure at the center. The folded-dipole pairs are in geometric progression in dimensions with a scaling factor $k$ from the most inner pairs to the most outer ones and cascaded one after another. The radiation arms at two sides of a panel are connected at the center with an edge-coupled microstrip line (twin lines above ground plane) which is excited differentially through two coaxial cables. The antenna is defined by 9 geometric parameters for each panel, with the definition shown in Fig.~\ref{fig:parameters} and listed in Table~\ref{tab:parameters} along with their corresponding optimal values.

The antenna is designed via optimization to produce maximum MIMO efficiency according to \eqref{eq:MIMO_efficiency} for both Random-LOS and RIMP and low reflection coefficient ($S_{11}$) over the frequency band from $1.6$ to $2.8$~GHz. In order to have wideband performance, the number of the cascaded folded-dipole pairs should be large, which consequently increases the size of the antenna. It is observed that 8 folded-dipole pairs are enough to achieve the bandwidth requirement. Then, all geometric parameters, according to Table~\ref{tab:parameters}, have gone through optimization of PoD and $S_{11}$ to be determined.

To find the initial values of the geometric parameters for the optimization, we tune all parameters one by one while keeping others fixed. The function of every parameter i.e. how it will affect the $S_{11}$ and PoD, can be observed through this process. Then, the initial values and the parameter scanning range have been determined.

\begin{table}[b]
	\centering
	\caption{Description and values of geometric parameters}
	\label{tab:parameters}
	\begin{tabular}{lp{50mm}p{7mm}p{7mm}}
		\hline
		Par.      & Description & Low branch & High branch\\ \hline
		$k$              & scaling factor &1.2739&1.2757\\
		$k_d$           & scaling factor of horizontal distance between dipole pairs&0.7218&0.7661\\
		$k_{d_a}$    & scaling factor of dipole width&0.0097&0.0093\\
		$k_{d_c}$    & scaling factor of gap width in dipole pair&0.0349&0.0283\\
		$k_{w_a}$    & scaling factor of dipole arm width&0.0213&0.0216\\
		$k_l$           & scaling factor of dipole arm length&0.2137&0.2066\\
		$s$              & separation between transmission lines [mm]&3.1613&2.9210\\
		$w_t$          & width of the transmission line [mm]&1.2999&1.3149\\ \hline
	\end{tabular}
\end{table}

\subsection{Optimization}
Genetic Algorithm \cite{rahmat1999electromagnetic} is used for optimization process. The geometric parameters influence the performance of both $S_{11}$ and PoD, and thus are treated as genes. Each generation consists of 400 samples. Initial population is randomly generated with a uniform distribution within the range which was previously obtained through parameter sweep process.

The population is ranked according to the value of their maximum $S_{11}$ over the frequency band. The PoD of all the samples is also evaluated via ViRM-Lab through the simulated far-field function. Fifteen samples of the four hundred with the lowest maximum $S_{11}$ and acceptable PoD are selected as elites and given the chance to mate each other pairwise. Each pair will generate 2 children. Then, the roulette wheel selection rule is used to pick fifteen more samples in the remaining 385 samples to generate offsprings in the same way as the elites. The optimization converges after only 3 generations.

\begin{figure}[t]
	\centering
	\includegraphics[width=0.7\columnwidth]{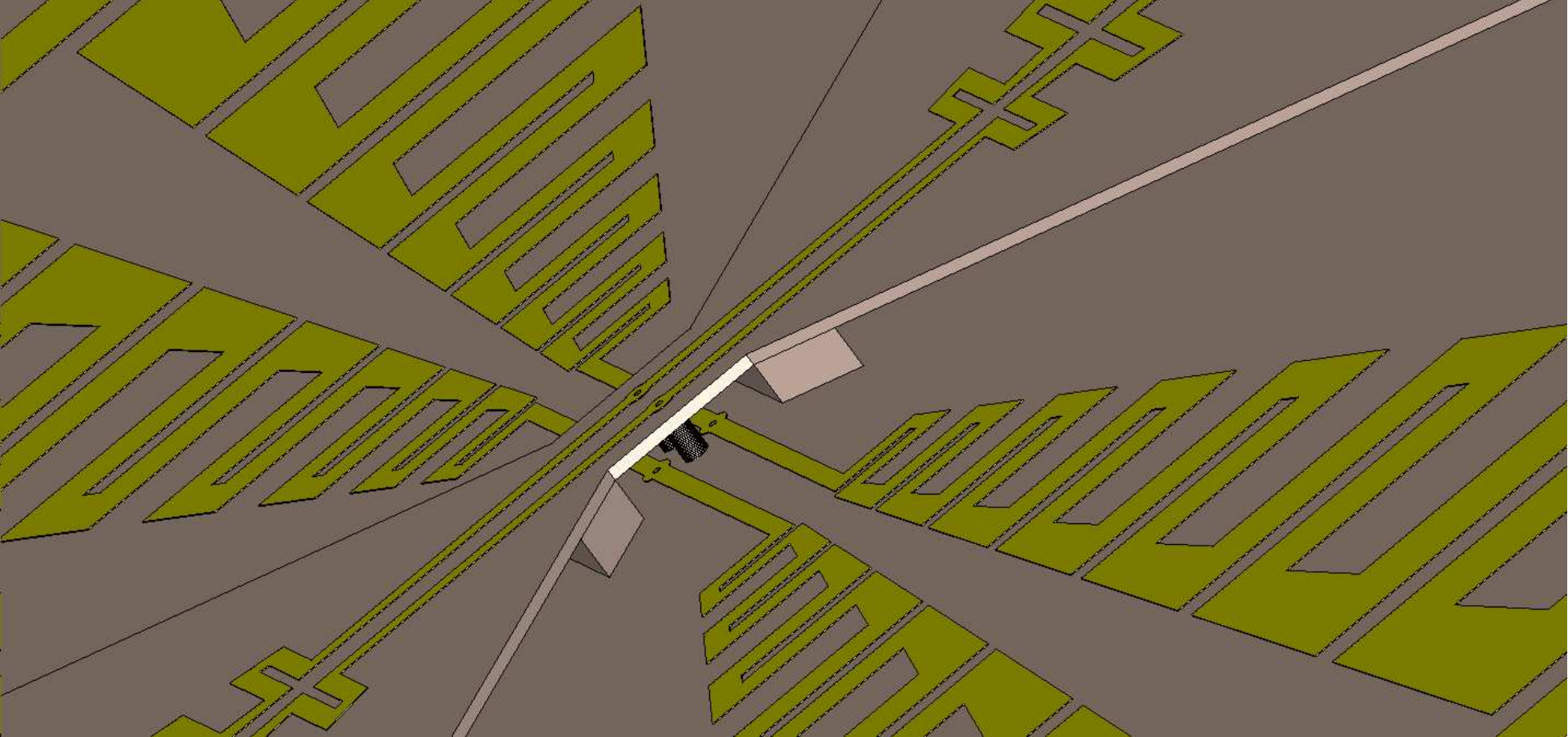}\\
	(a) Transmission line part\\
	\includegraphics[width=0.7\columnwidth]{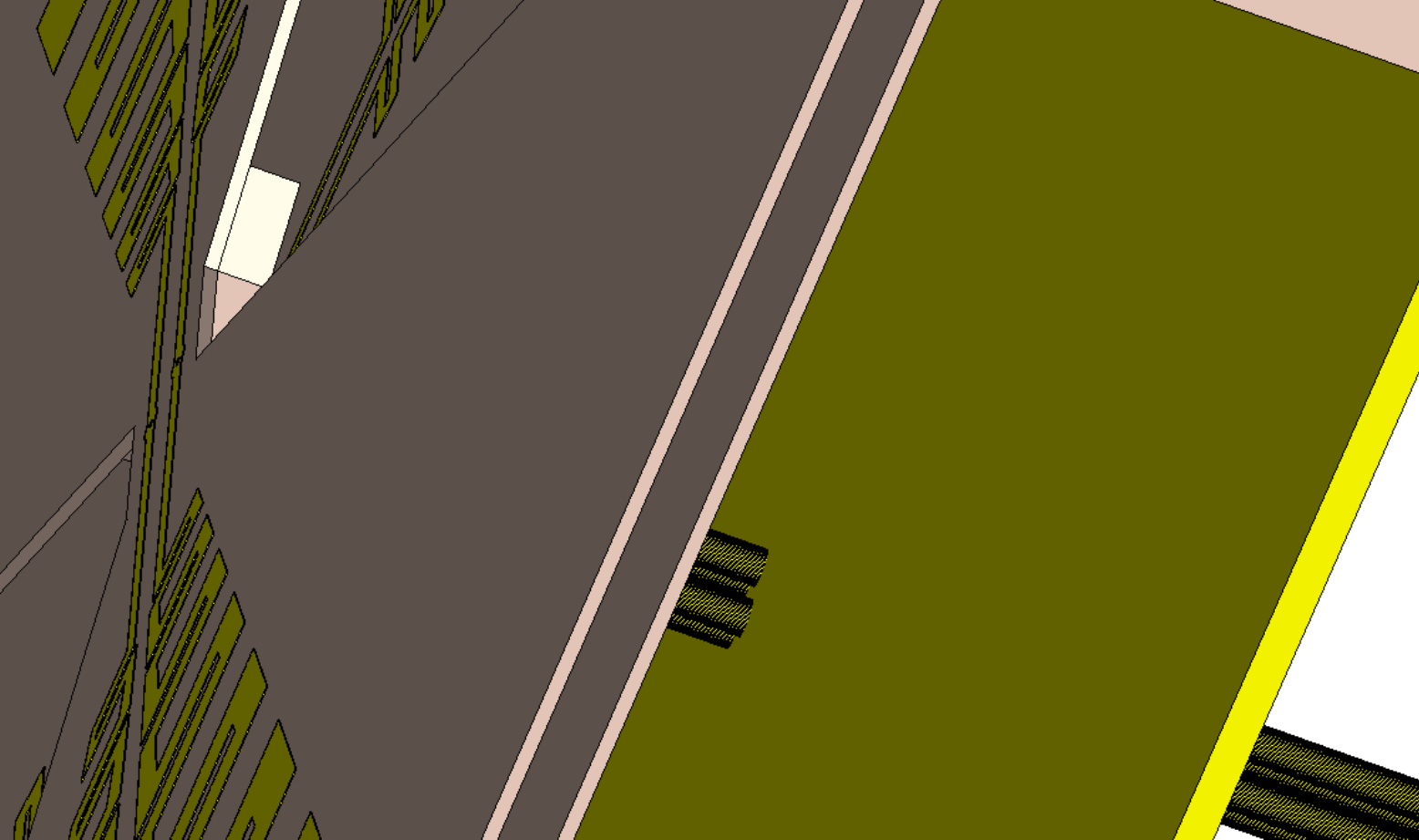}\\
	(b) Side view\\
	\caption{Detail of the CST model of the planar Eleven MIMO antenna.}
	\label{fig:DetailCSTmodel}
\end{figure}

The total size of the antenna is $337\times337\times37 \text{mm}^3$ (see Fig.~\ref{fig:CSTmodel} and details in Fig.~\ref{fig:DetailCSTmodel}).

\section{Simulation and Measurement Results}
\begin{figure}[t]
	\begin{center}
		\includegraphics[width=0.5\columnwidth]{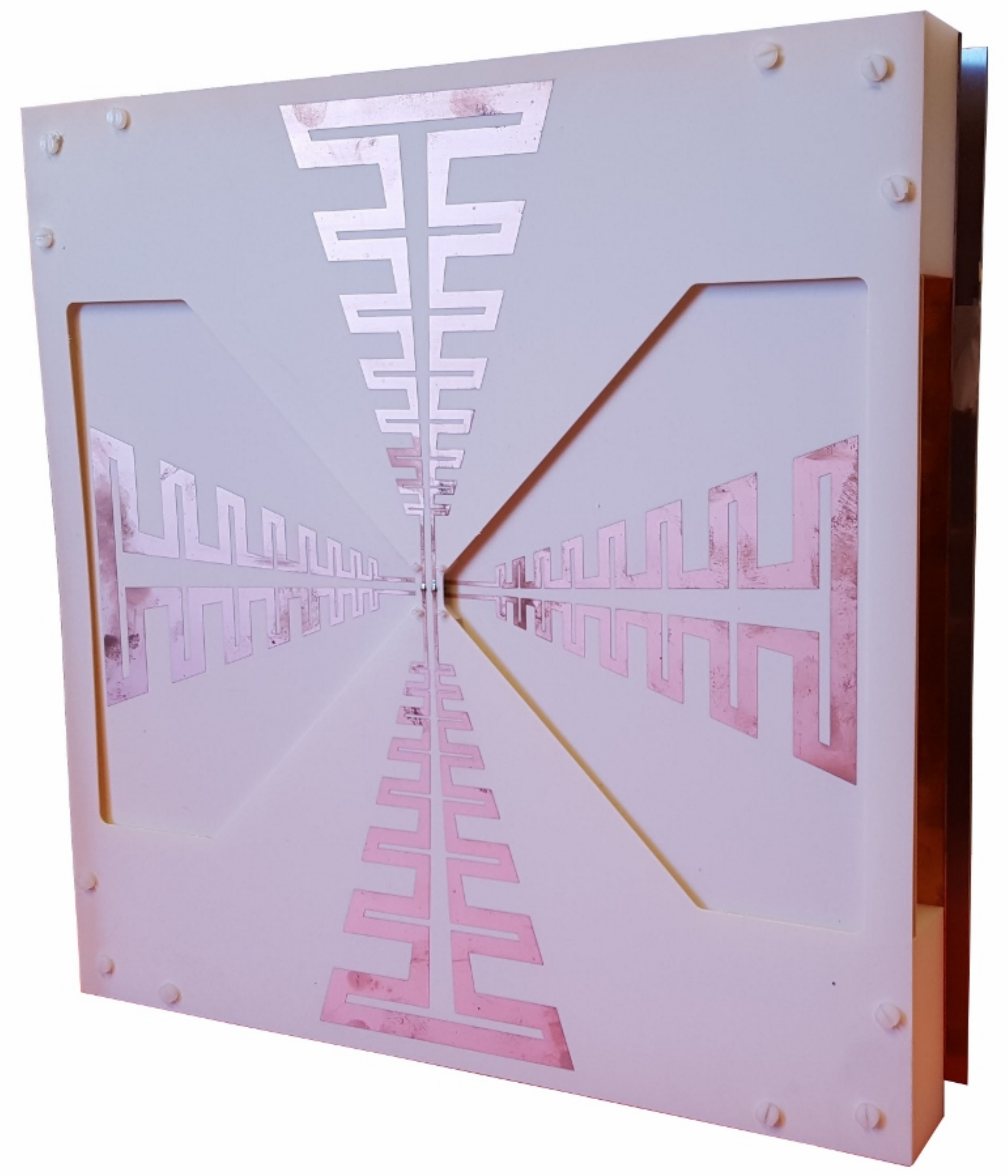}
		\caption{The fabricated prototype of the planar Eleven antenna.}
		\label{fig:prototype}
	\end{center}
\end{figure}

Fig.~\ref{fig:prototype} shows the fabricated prototype of the optimized planar Eleven antenna. The prototype is made for 4-port dual polarization. By using a wideband hybrid junction as a balun, the prototype is operated in the 2-port mode, and all measurements were done for this 2-port antenna. The configuration of this antenna makes it very flexible to also have 8 ports with minor modification of the feeding structure. We present also the simulation results of this Eleven MIMO antenna with 4 ports and 8 ports.

\begin{figure}[!t]
	\begin{tabular}{cc}
		\includegraphics[width=0.45\columnwidth]{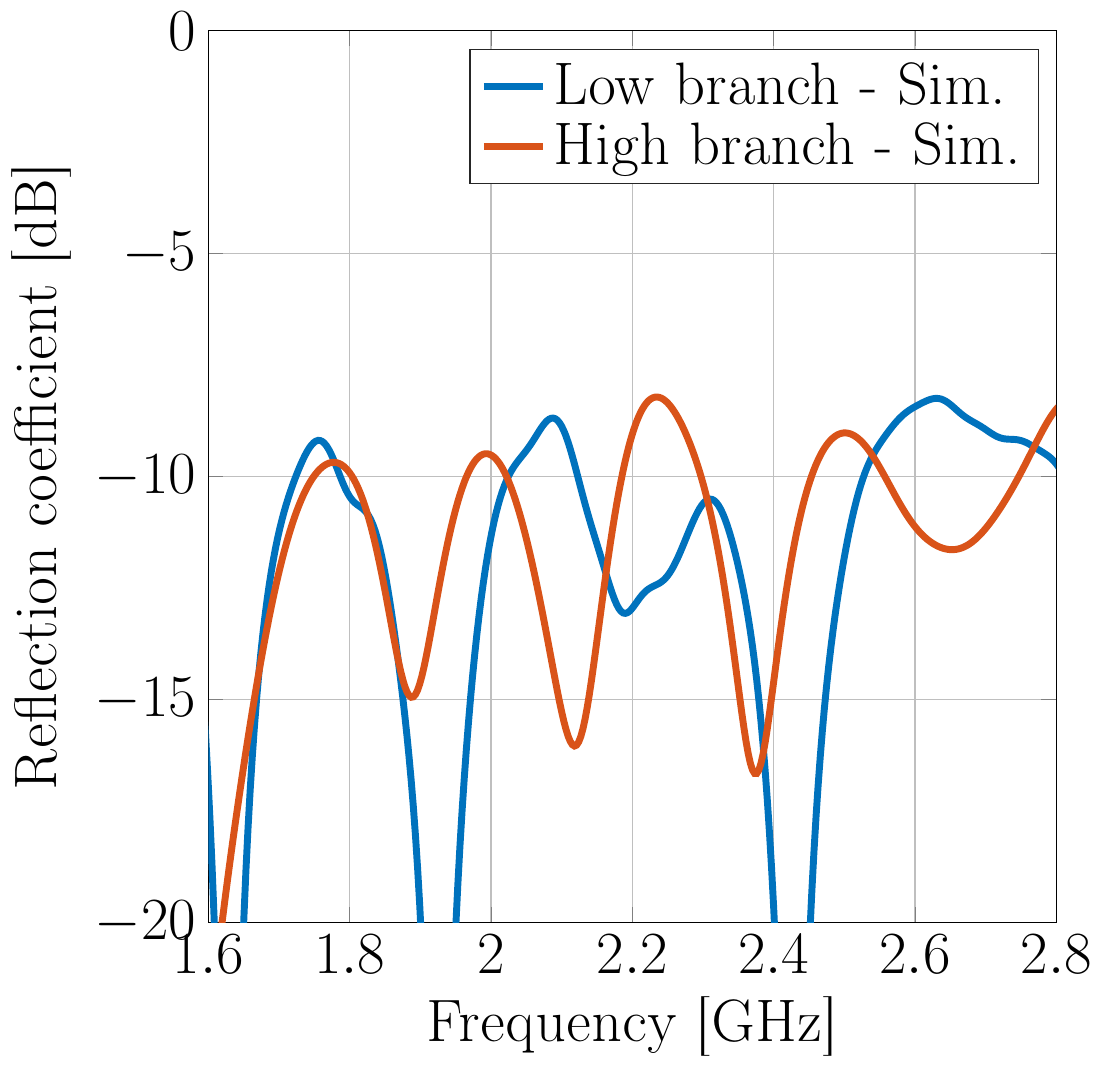}&
		\includegraphics[width=0.44\columnwidth]{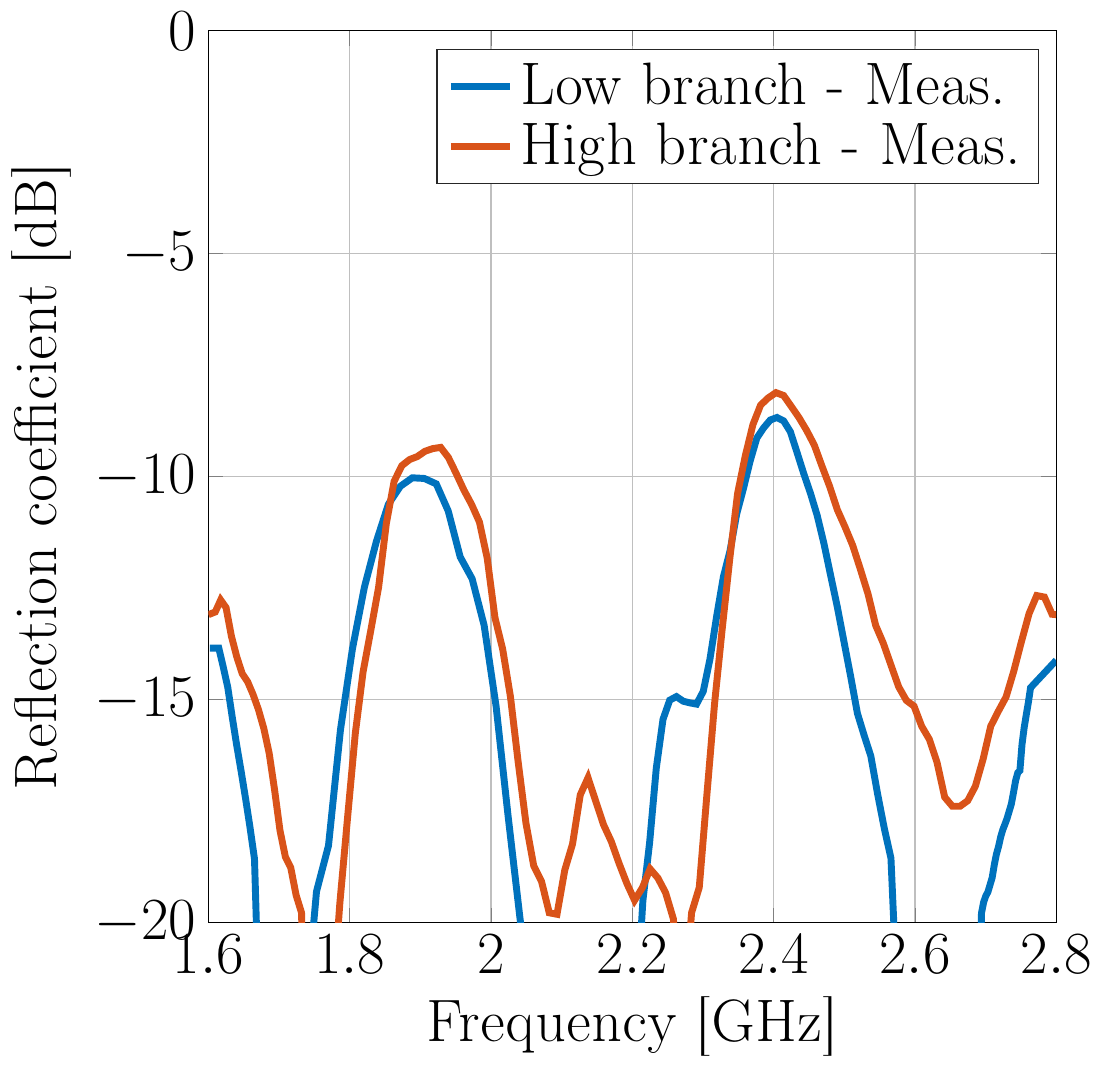}\\
		(a) & (b)
	\end{tabular}
	\caption{The reflection coefficient (a) simulated with differential feeding, and (b) measured with wideband hybrid junction.}
	\label{fig:S11}
\end{figure}

Fig.~\ref{fig:S11} shows the simulated and measured reflection coefficient of the 2-port dual polar antenna. Both the simulation model and the prototype of the 2-port antenna require a balun for feeding the antenna. The 2-port antenna mode is simulated by using ideal differential feeding in CST. For the prototype a wideband hybrid junction is used as the balun. Due to this reason, the actual simulated and measured reflection coefficients of the 2-port mode have a certain difference. The reflection coefficient is not included in the MIMO efficiency calculations. However, we need to keep in mind that in general poor matching will degrade the MIMO performance.

\subsection{Dual-Polarized 2-port MIMO Antenna}
\begin{figure}[!t]
	\begin{center}
		\includegraphics[width=0.49\columnwidth]{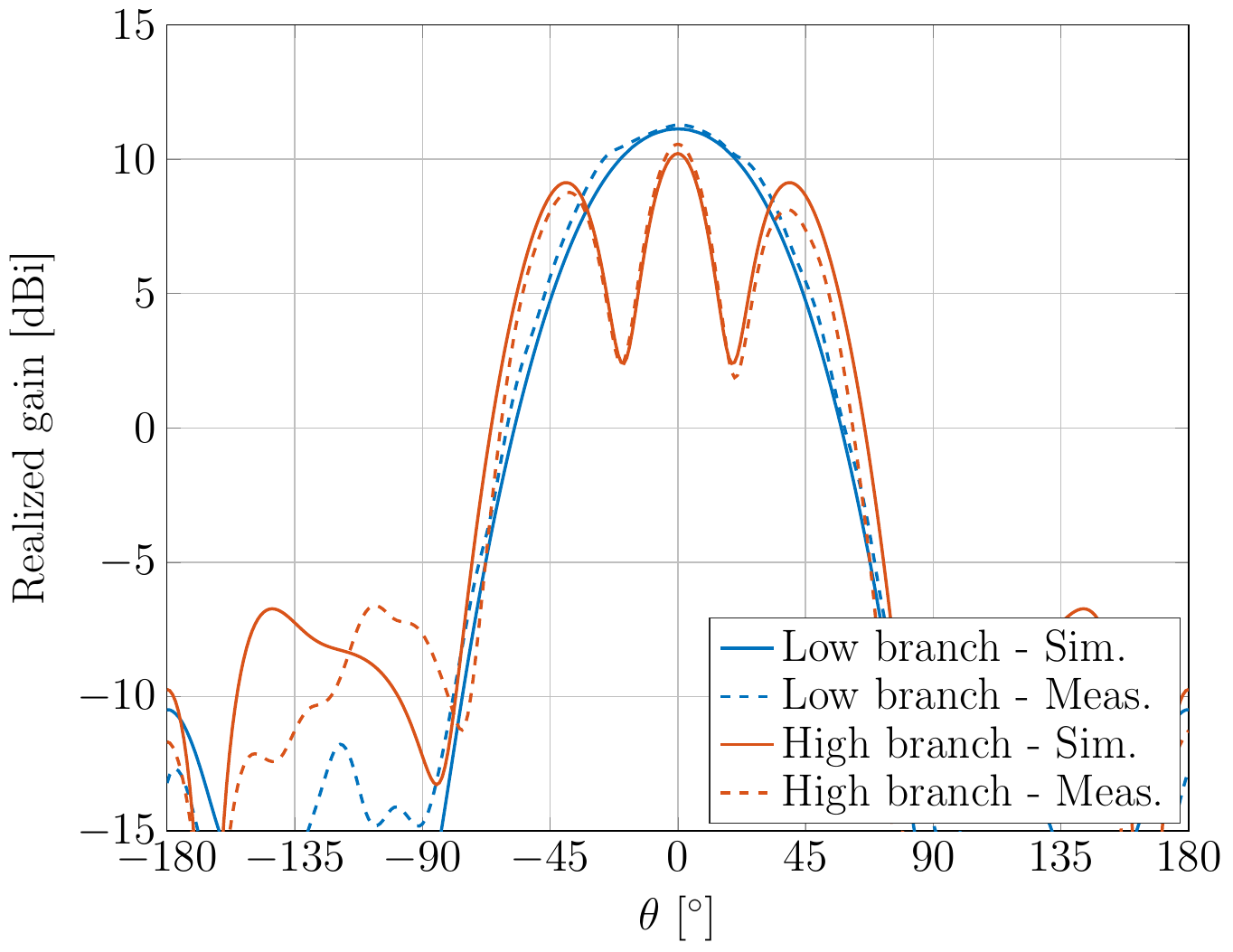}
		\includegraphics[width=0.49\columnwidth]{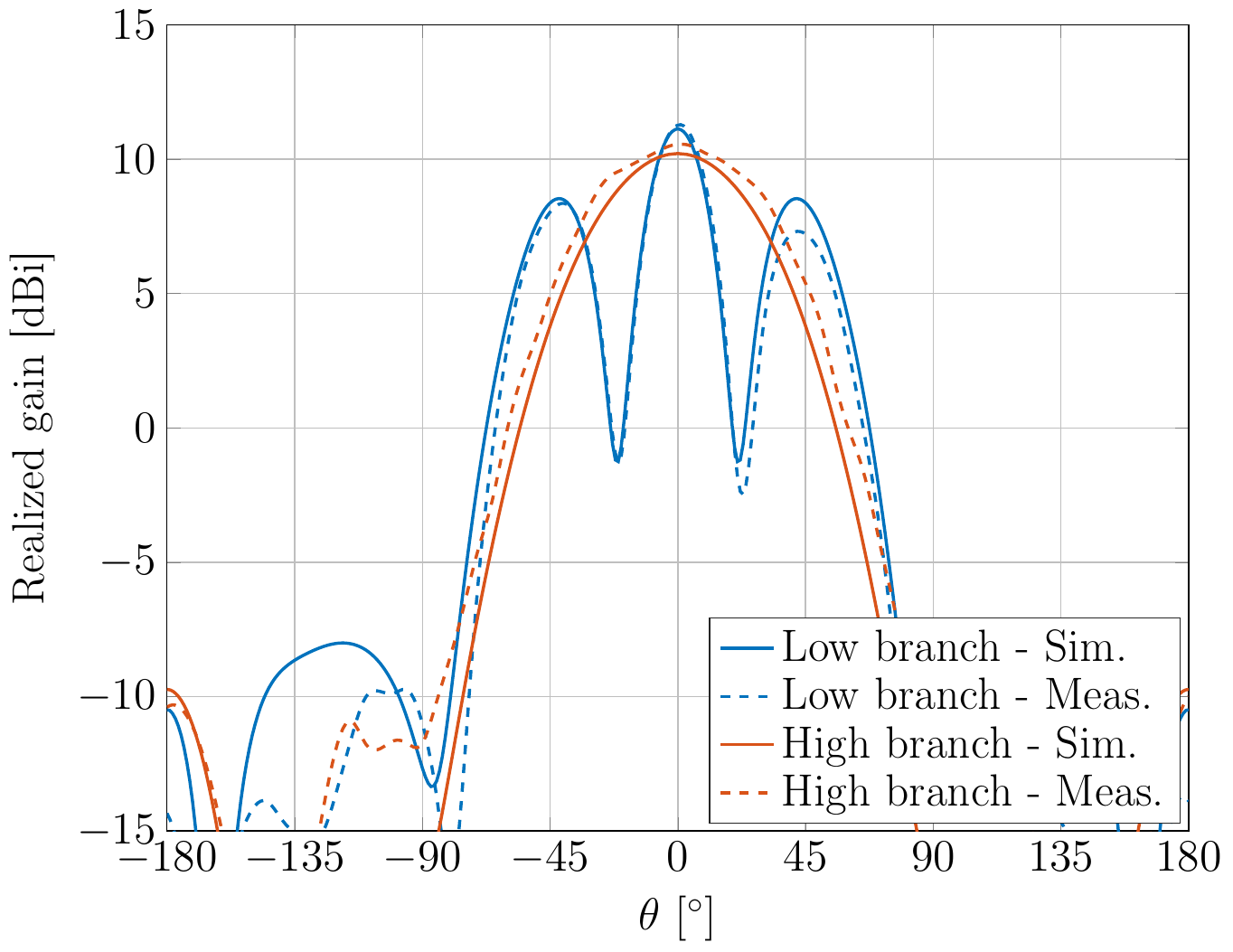}
	\end{center}\vspace{-4mm}
	\small\hspace{22mm}(a)\hspace{41mm}(b)\vspace{-3mm}
	\begin{center}
		\includegraphics[width=0.49\columnwidth]{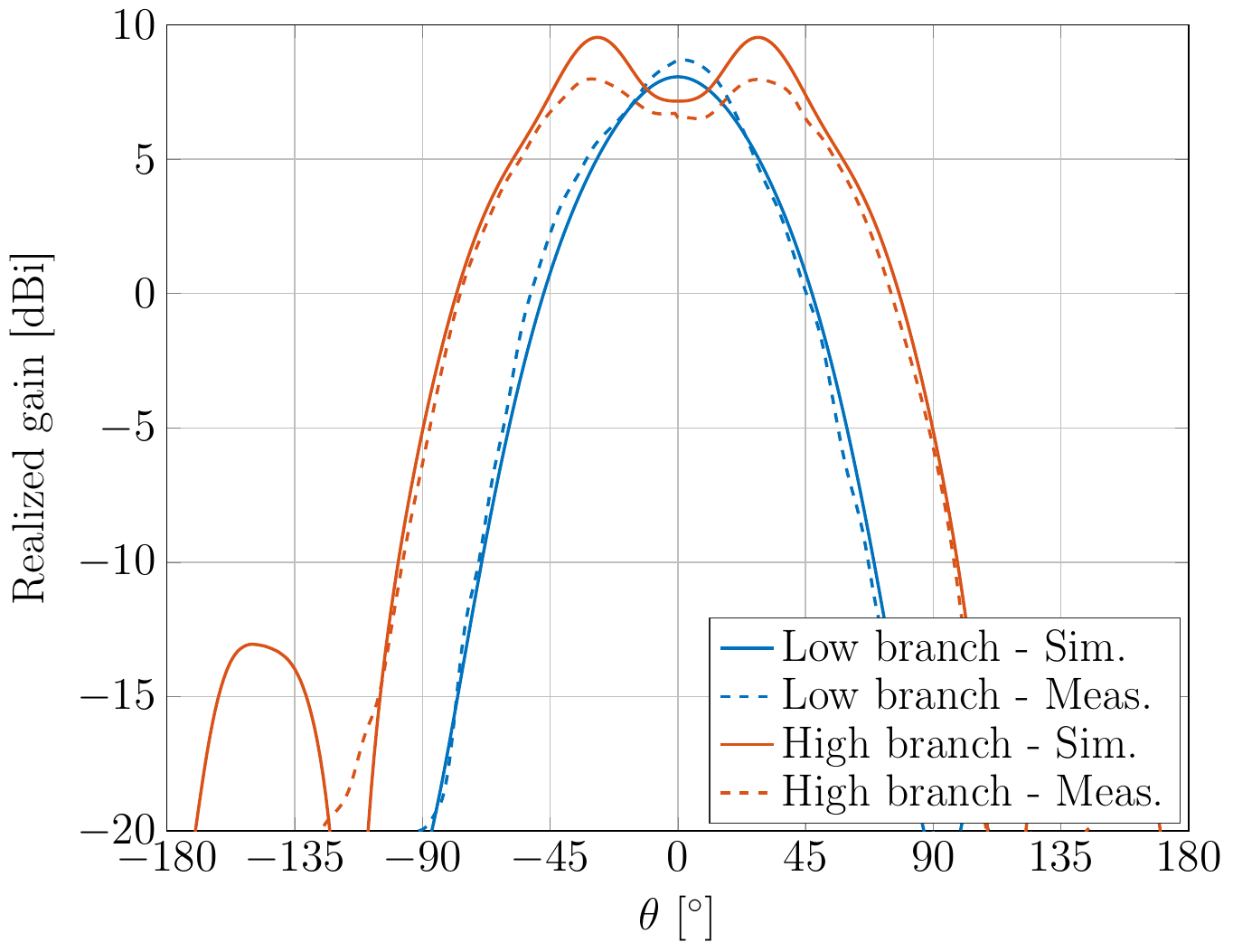}
		\includegraphics[width=0.49\columnwidth]{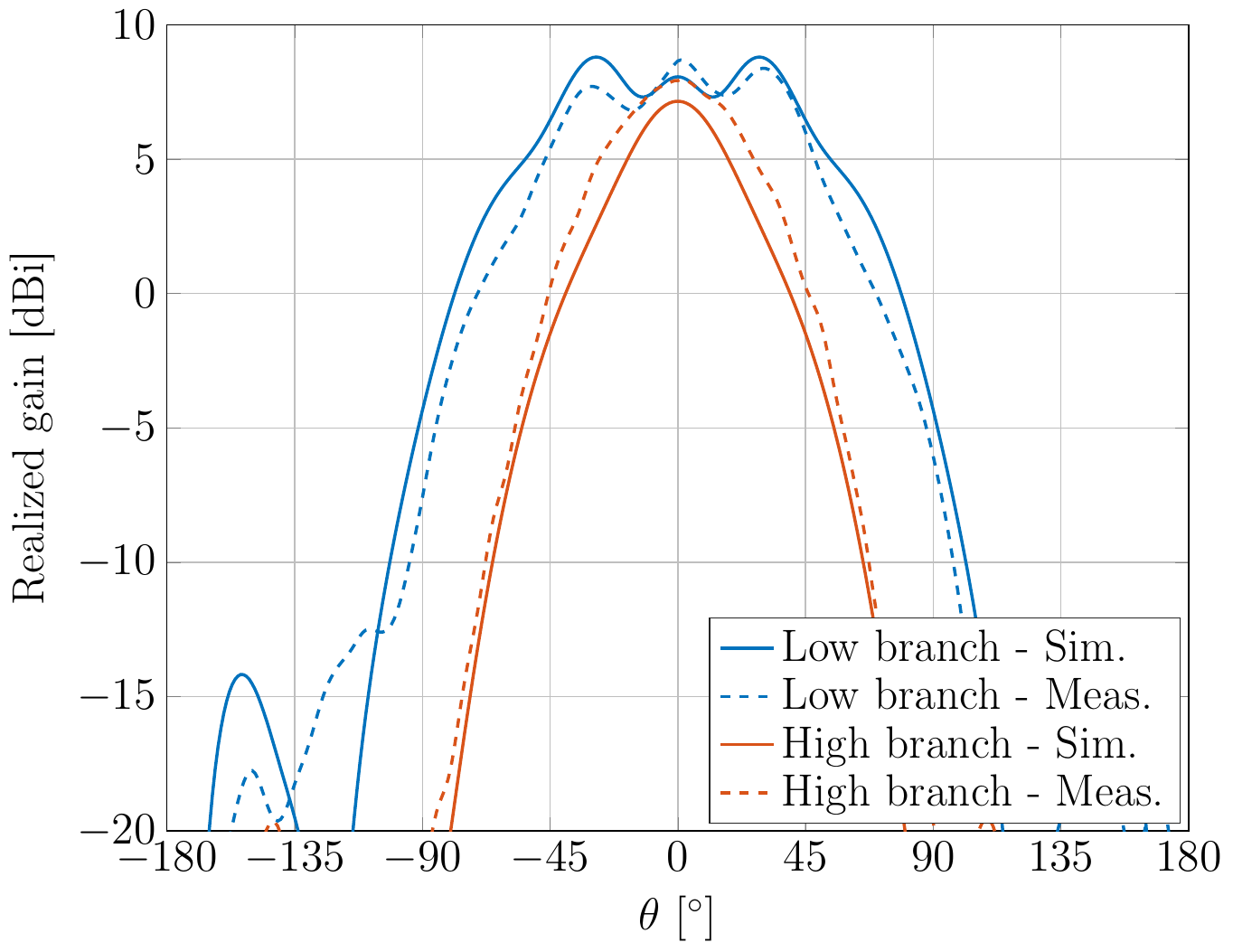}
	\end{center}\vspace{-4mm}
	\small\hspace{22mm}(c)\hspace{41mm}(d)\vspace{-2mm}
	\begin{center}
		\includegraphics[width=0.49\columnwidth]{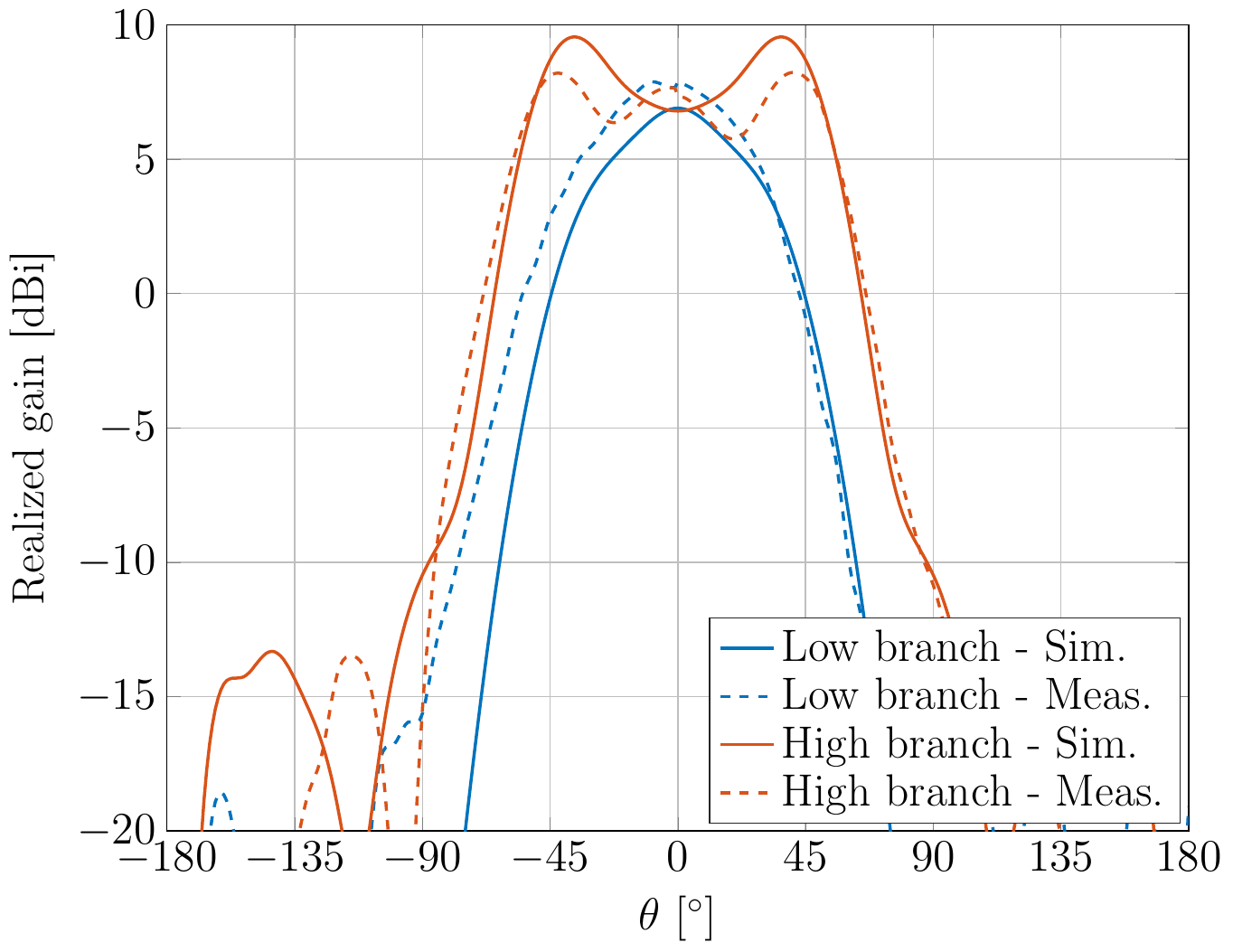}
		\includegraphics[width=0.49\columnwidth]{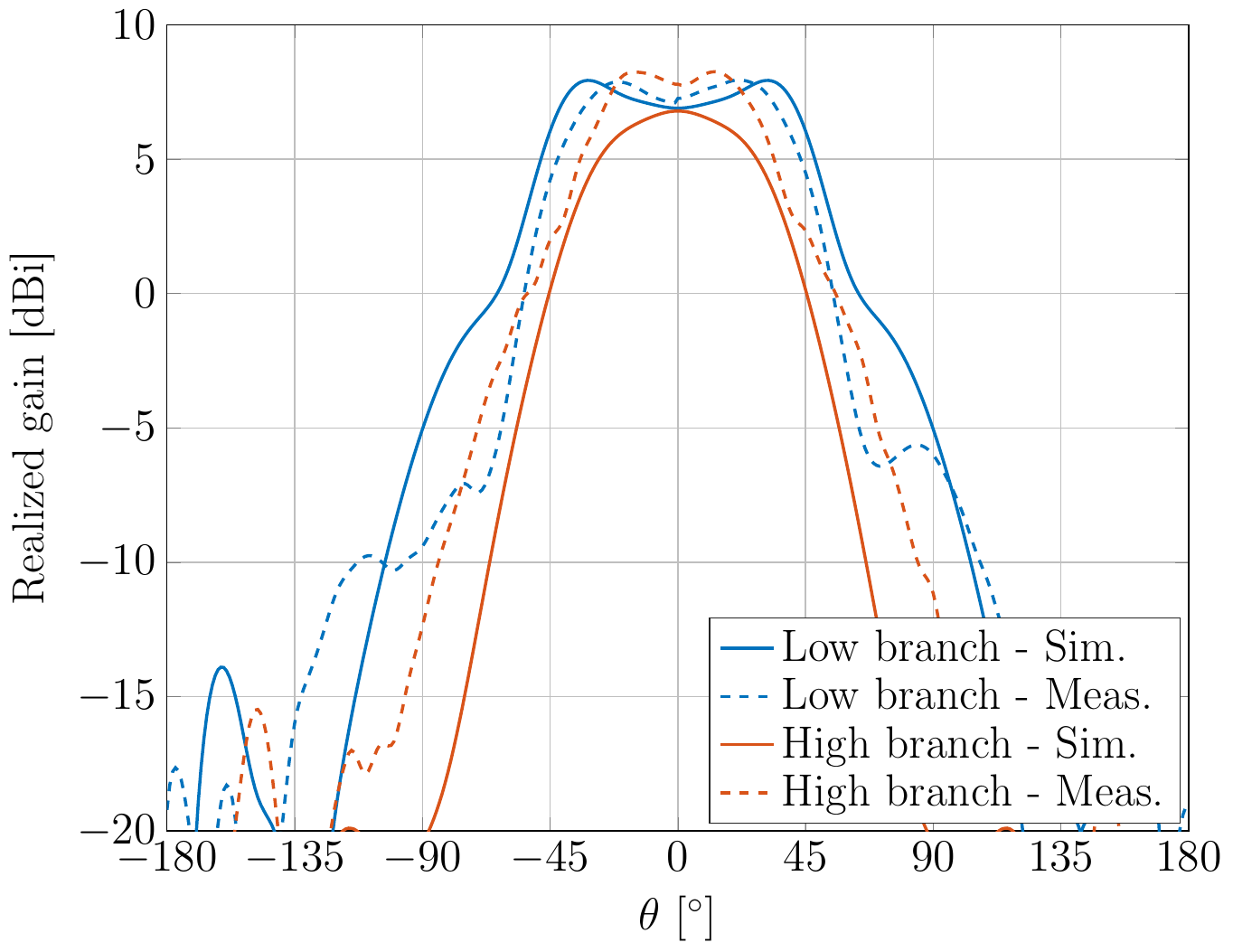}
	\end{center}\vspace{-4mm}
	\small\hspace{22mm}(e)\hspace{41mm}(f)\vspace{-3mm}
	\caption{\small Simulated and measured radiation patterns of the 2-port antenna at (top) 1.6 GHz, (center) 2.2 GHz, and (bottom) 2.8 GHz, in (left) $\phi=0$ and (right) $\phi=90^\circ$ planes. Here, the high branch is oriented along $x$-axis and the low branch along $y$-axis.}
	\label{fig:pattern}
\end{figure}

Assuming the high branch along $x$-axis and the low branch along $y$-axis, the simulated and measured radiation patterns in $\phi=0$ and $\phi=90^\circ$ planes are plotted in Fig.~\ref{fig:pattern}, for the beginning, center and the end of the frequency band. We can observe that there is a good agreement between simulations and measurements. The far-field functions (both amplitude and phase) of the antenna for the two orthogonal polarizations have been measured for the angle steps of $\Delta\theta=5^\circ$ and $\Delta\phi=15^\circ$. Then, interpolation was used to get the measured far-field function with angle steps of $\Delta\theta=1^\circ$ and $\Delta\phi=1^\circ$. The simulated far-field function was obtained from CST with the same resolution.

\begin{figure}[!t]
	\begin{center}
		\includegraphics[width=0.8\columnwidth]{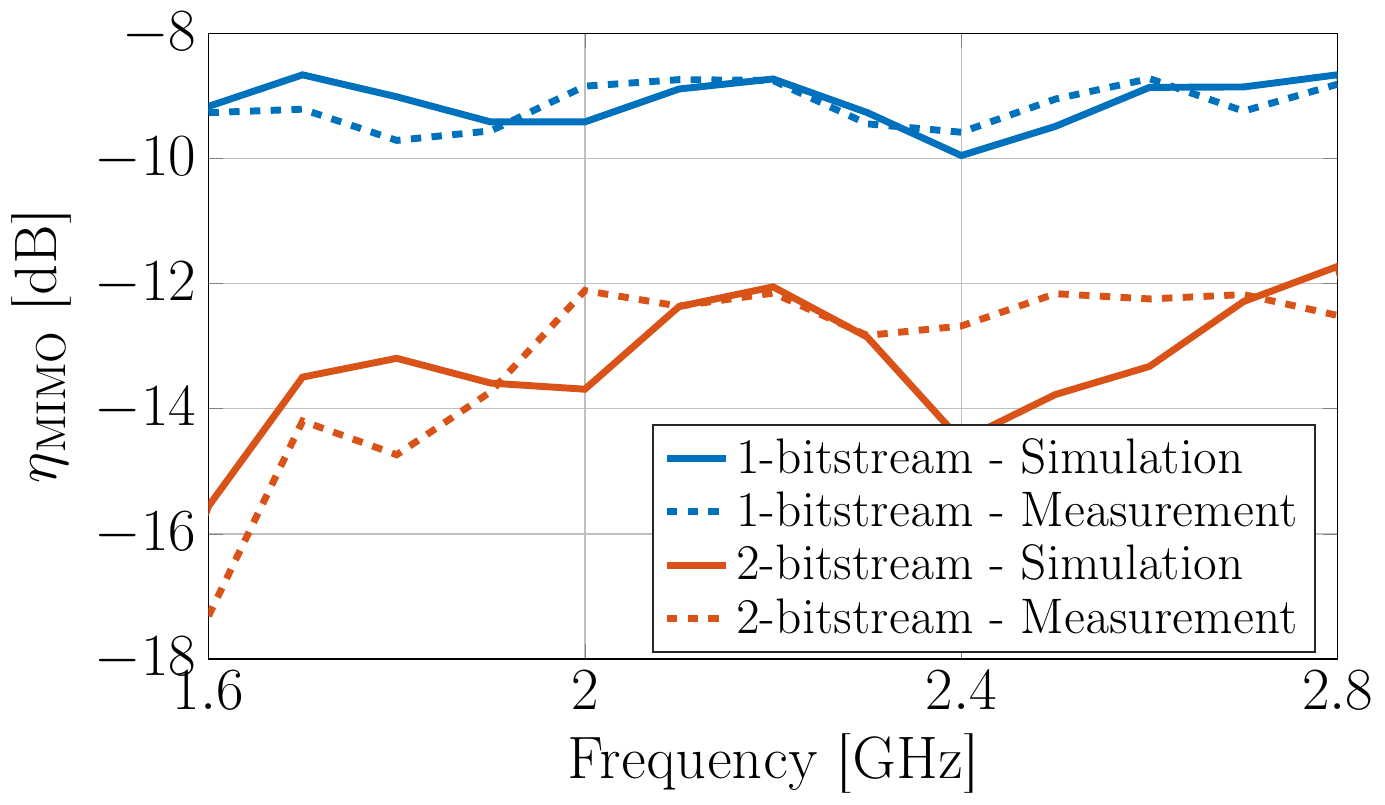}
	\end{center}
	\caption{\small 1-bitstream and 2-bitstream MIMO efficiency of the two port antenna in Random-LOS.}
	\label{fig:efficiency_2ports_rlos}
\end{figure}

\begin{figure}[!t]
	\begin{center}
		\includegraphics[width=0.8\columnwidth]{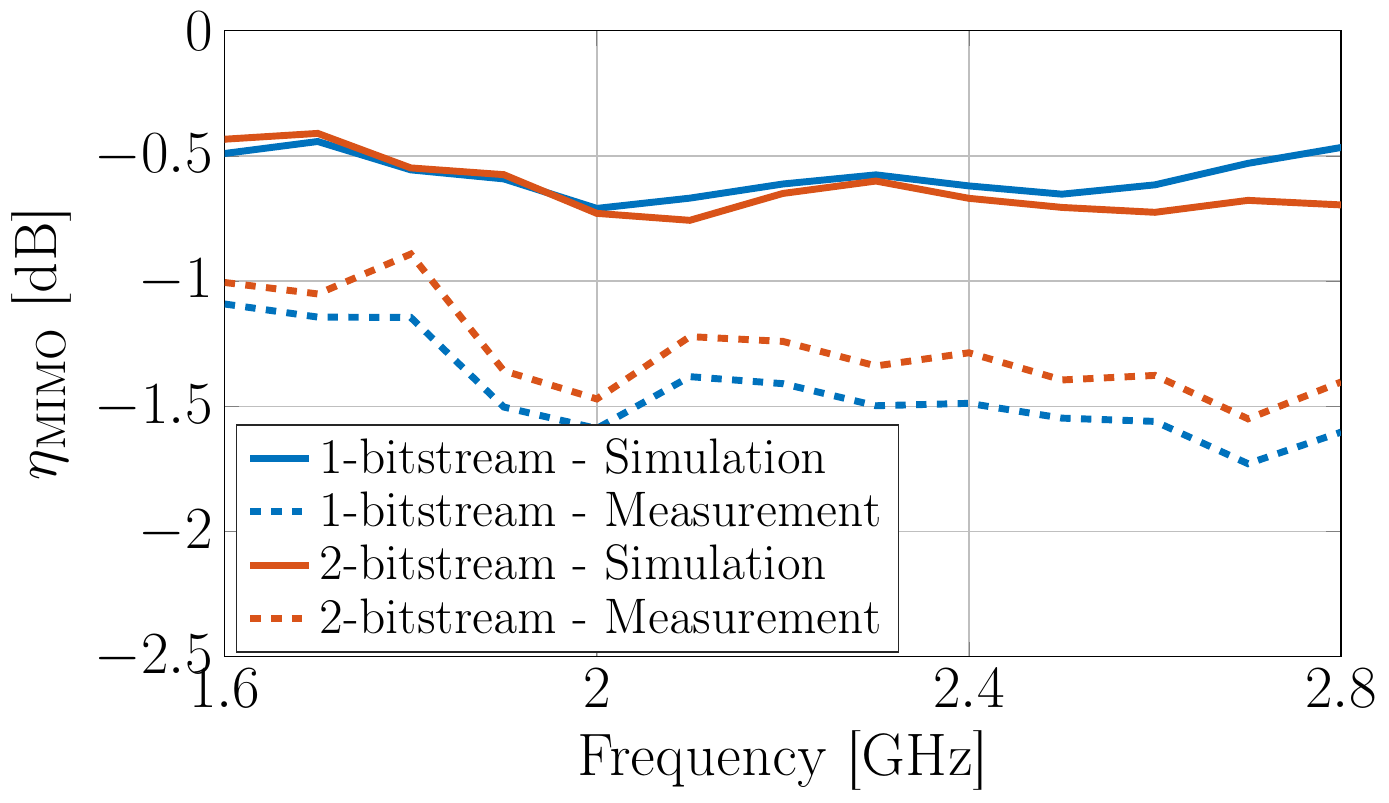}
	\end{center}
	\caption{\small 1-bitstream and 2-bitstream MIMO efficiency of the two port antenna in RIMP.}
	\label{fig:efficiency_2ports_rimp}
\end{figure}

The MIMO efficiency of the 2-port antenna for both 1-bitstream $2\times 1$ and 2-bitstream $2\times 2$ systems based on the simulated and the measured far-field functions is plotted in Fig.~\ref{fig:efficiency_2ports_rlos} and Fig.~\ref{fig:efficiency_2ports_rimp} for Random-LOS and RIMP, respectively. These figures show good agreement between the simulation and measurement results. Furthermore we can observe that the MIMO efficiency has relatively little variation over the frequency bandwidth.

\begin{figure*}[t!]
	\begin{center}
		\includegraphics[width=0.3\textwidth]{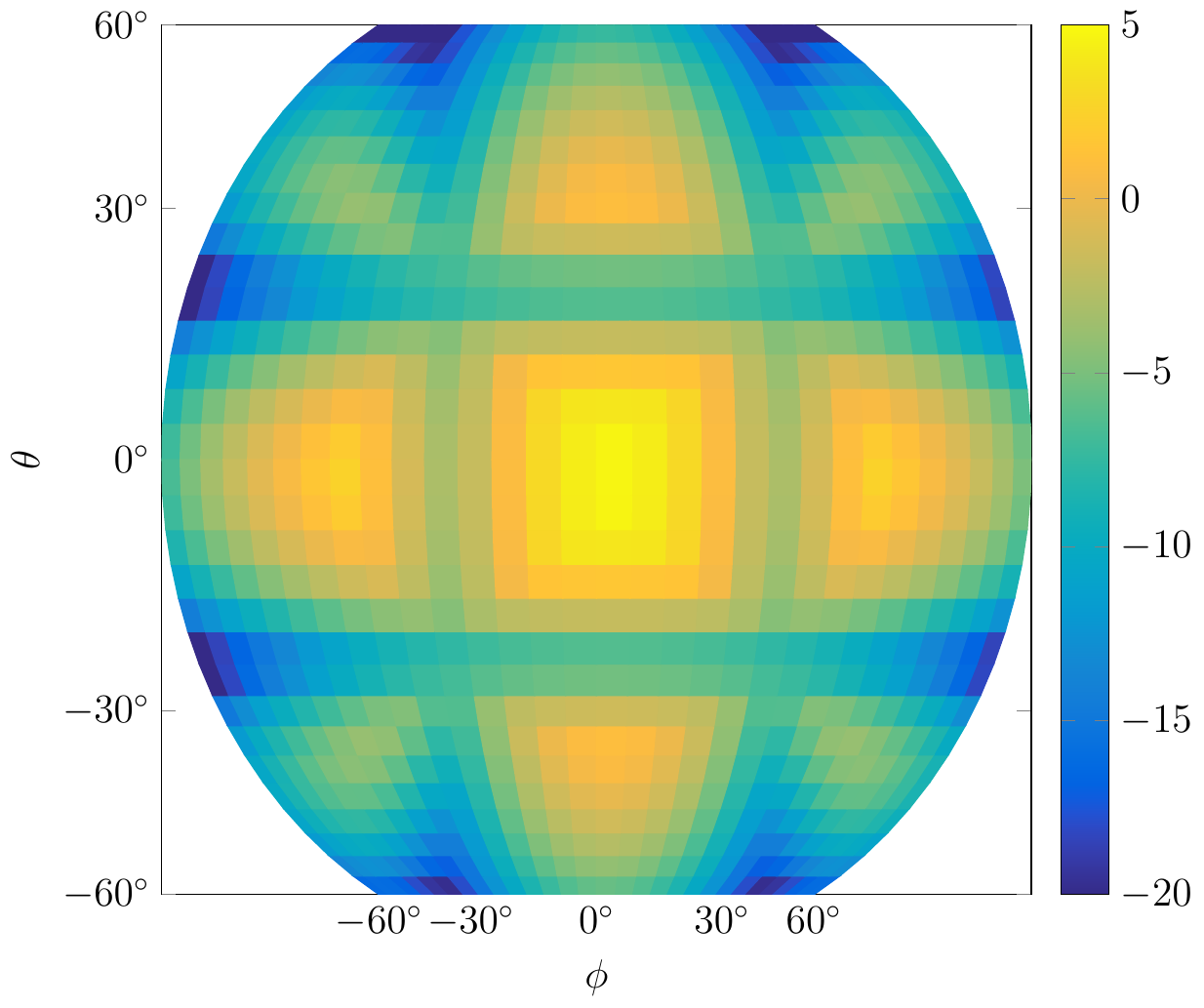}
		\includegraphics[width=0.3\textwidth]{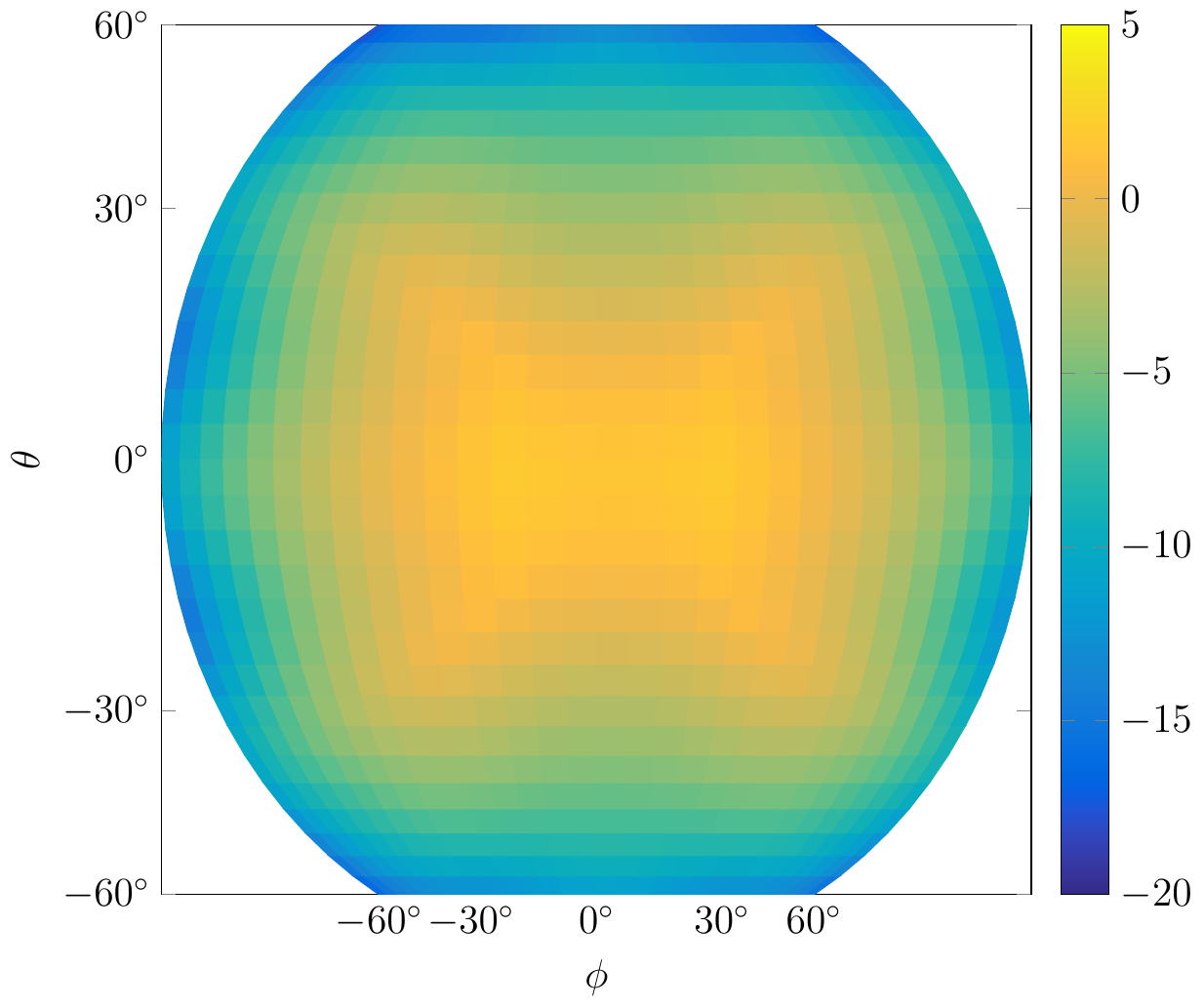}		
		\includegraphics[width=0.3\textwidth]{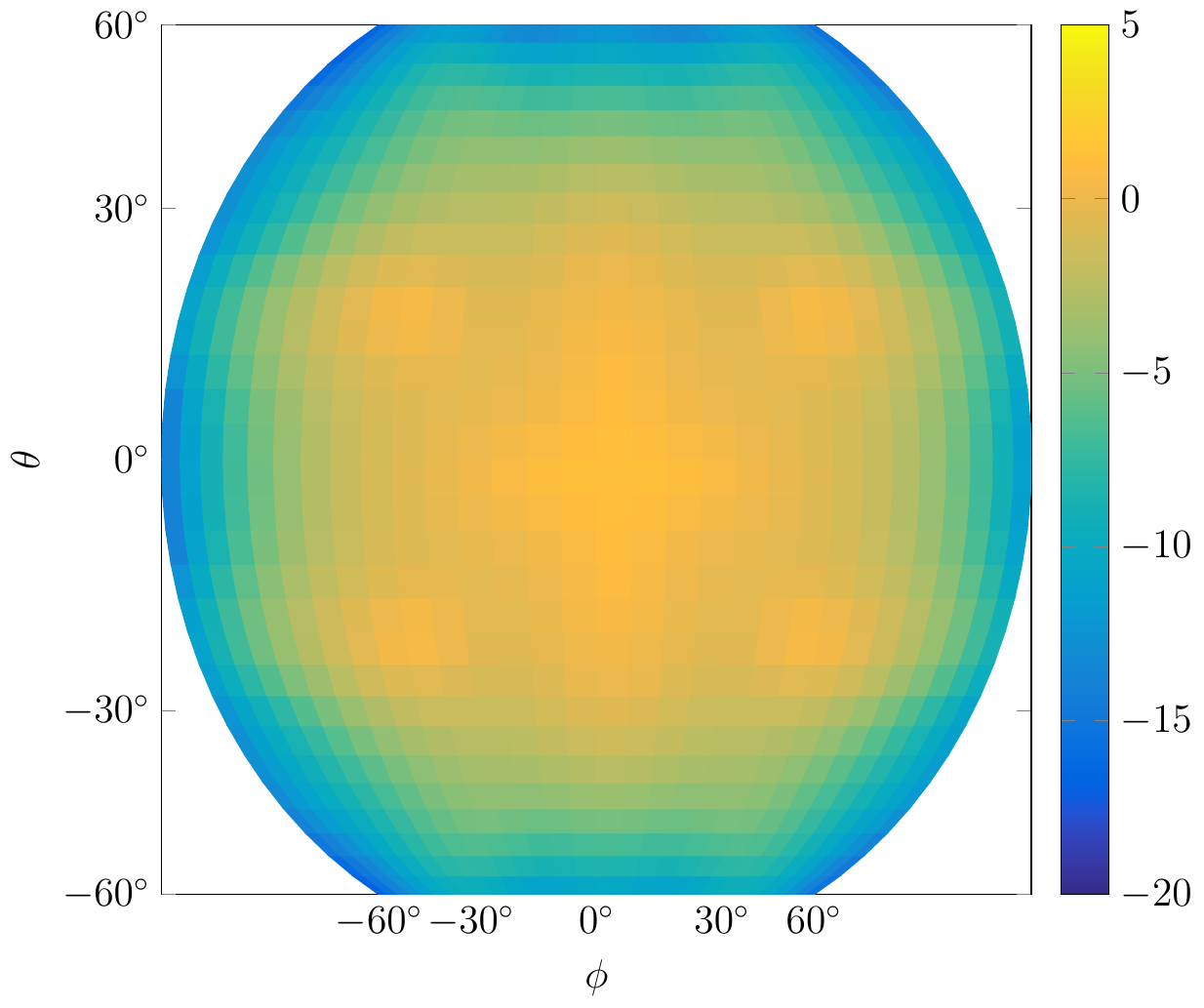}		
	\end{center}\vspace{-4mm}
	\small\hspace{33mm}(a)\hspace{52mm}(b)\hspace{52mm}(c)\\ \vspace{-6mm}
	\caption{\small 2-bitstream MIMO coverage plots of the two port antenna in Random-LOS at (a) 1.6 GHz, (b) 2.2 GHz, and (c) 2.8 GHz.}
	\label{fig:coverage}
\end{figure*}

The Random-LOS MIMO efficiency is low at lower frequencies as observed in Fig.~\ref{fig:efficiency_2ports_rlos}. In order to gain better insight into the reason for this low efficiency, we should look at the spatial distribution of MIMO efficiency in MIMO coverage plots. MIMO efficiency defined by \eqref{eq:MIMO_efficiency} can be calculated for individual AoAs, where only the polarization is random. This is specially useful when dealing with 2-bitstream systems in Random-LOS. The reference antenna's coverage area is still the same as before. Therefore, at some AoAs the ratio in \eqref{eq:MIMO_efficiency} can be larger than 1. The coverage plots of the planar Eleven antenna for 2-bitstream are shown in Fig.~\ref{fig:coverage} at three frequencies. It can be observed that at lower frequencies, the efficiency is very low at some directions. Whereas it is more homogeneous at higher frequencies. Comparing this to the plots in Fig.~\ref{fig:efficiency_2ports_rlos} we can observe how this corresponds to lower 2-bitstream MIMO efficiency at lower frequencies.

Dual-polarized antennas provide orthogonal and equal-amplitude far-field patterns only in limited directions. It is well known that with orthogonal and amplitude balanced antenna ports, it is possible to combine the channels on the ports of a dual-polarized antenna in such a way that the polarization mismatch can be completely compensated between the receiver and transmitter ends of the link \cite[Sec. 3.10]{kildal2015foundations}. But, the presence of polarization deficiencies between the two ports of the antenna, impairs this capability and leads to degradation in MIMO efficiency.

Assuming that the far-field functions of the two receiving antenna ports are defined as $\PhysVec{G}_1\left(\theta,\phi\right)$ and $\PhysVec{G}_2\left(\theta,\phi\right)$ at any direction ($\theta,\phi$) in space, two types of polarization deficiency, namely \emph{amplitude imbalance} ($I_\text{a}$) and \emph{polarization non-orthogonality} ($I_\text{p}$) are defined in \cite{razavi2016characterizing} as:
\begin{align}
\label{eq:ampimb}
I_\text{a}\left(\theta,\phi\right) &= \frac{\max\left\{|\PhysVec{G}_1|,|\PhysVec{G}_2|\right\}}{\min\left\{|\PhysVec{G}_1|,|\PhysVec{G}_2|\right\}}\\
I_\text{p}\left(\theta,\phi\right) &= \frac{\left|\PhysVec{G}_1\cdot\PhysVec{G}_2^*\right|}{\left|\PhysVec{G}_1\right|\left|\PhysVec{G}_2\right|}.
\end{align}
Both $I_\text{a}$ and $I_\text{p}$ are zero when the antenna ports are ideally orthogonal and amplitude-balanced, and they increase with the presence of polarization deficiencies.

\begin{figure}[!t]
	\begin{center}
		\includegraphics[width=0.49\columnwidth]{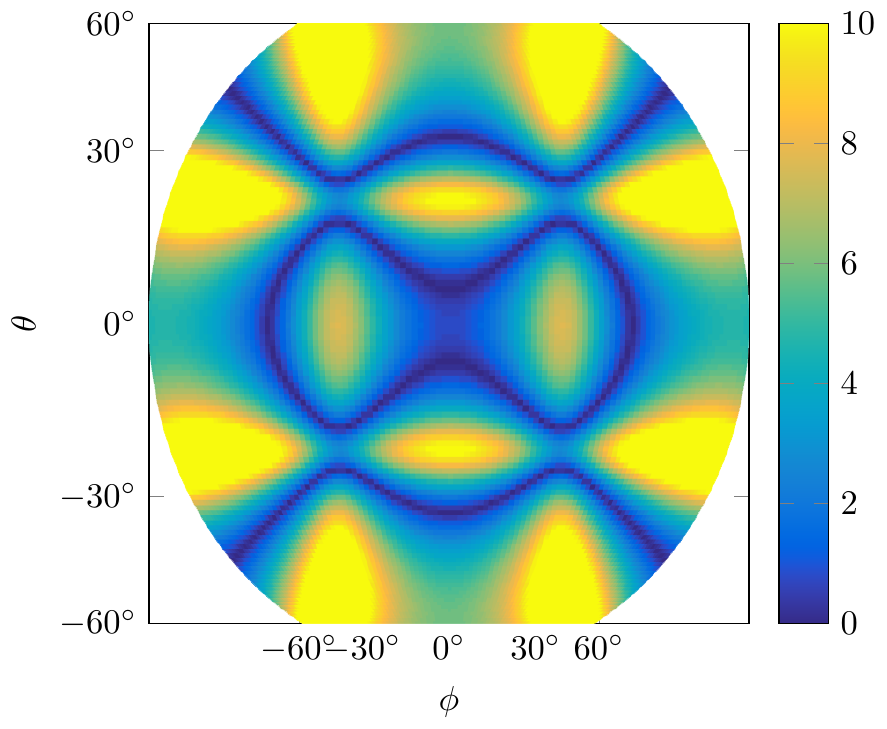}
		\includegraphics[width=0.49\columnwidth]{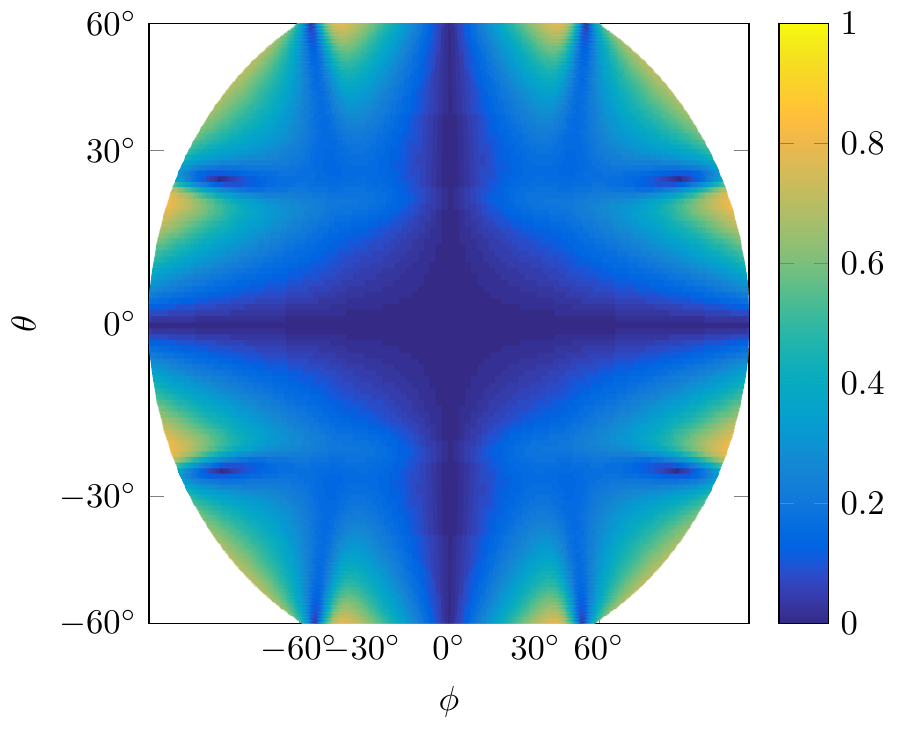}
	\end{center}\vspace{-4mm}
	\small\hspace{12mm}(a) $I_\text{a}$ at 1.6 GHz\hspace{22mm}(b) $I_\text{p}$ at 1.6 GHz\vspace{-3mm}
	\begin{center}
		\includegraphics[width=0.49\columnwidth]{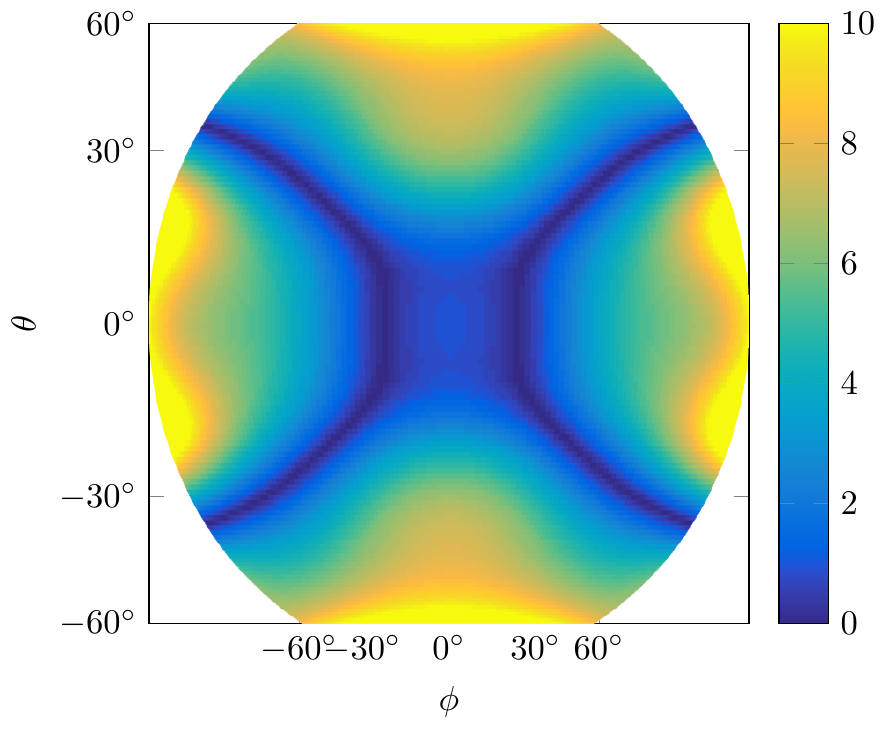}
		\includegraphics[width=0.49\columnwidth]{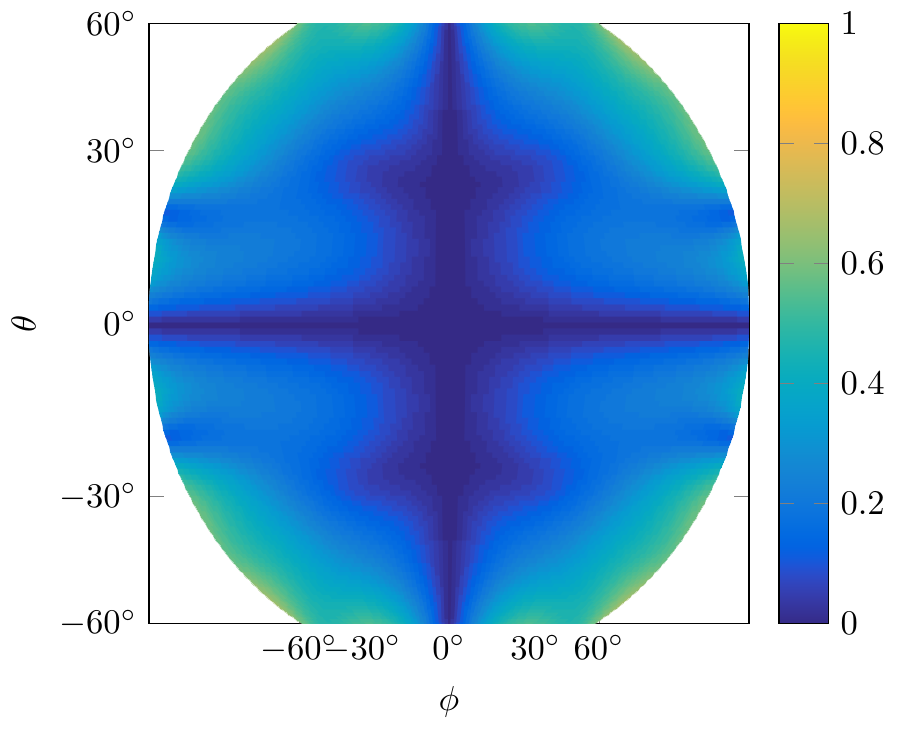}
	\end{center}\vspace{-4mm}
	\small\hspace{12mm}(c) $I_\text{a}$ at 2.2 GHz\hspace{22mm}(d) $I_\text{p}$ at 2.2 GHz\vspace{-3mm}
	\begin{center}
		\includegraphics[width=0.49\columnwidth]{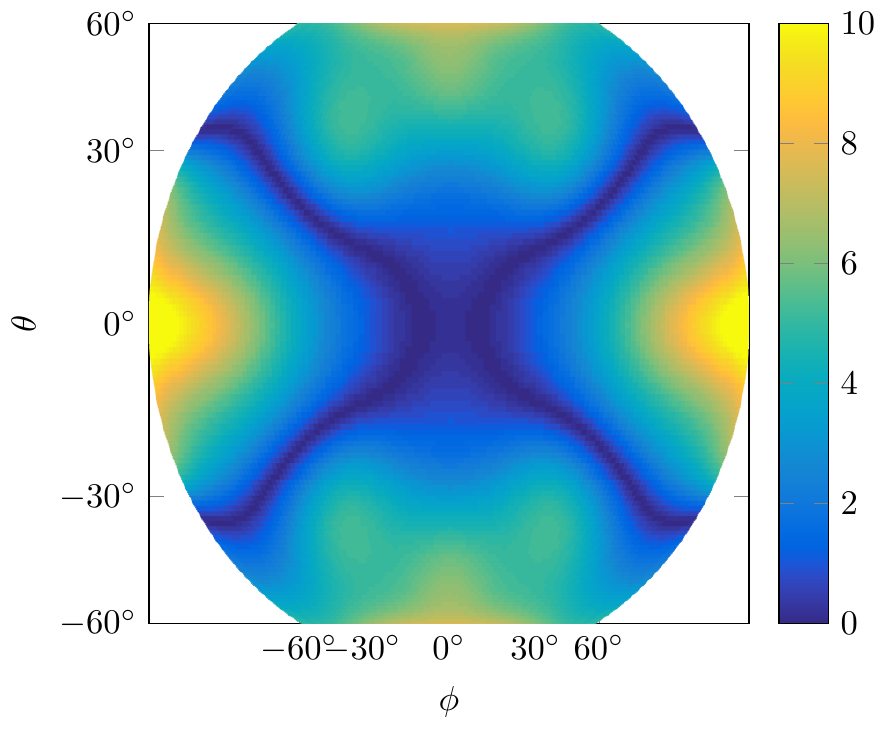}
		\includegraphics[width=0.49\columnwidth]{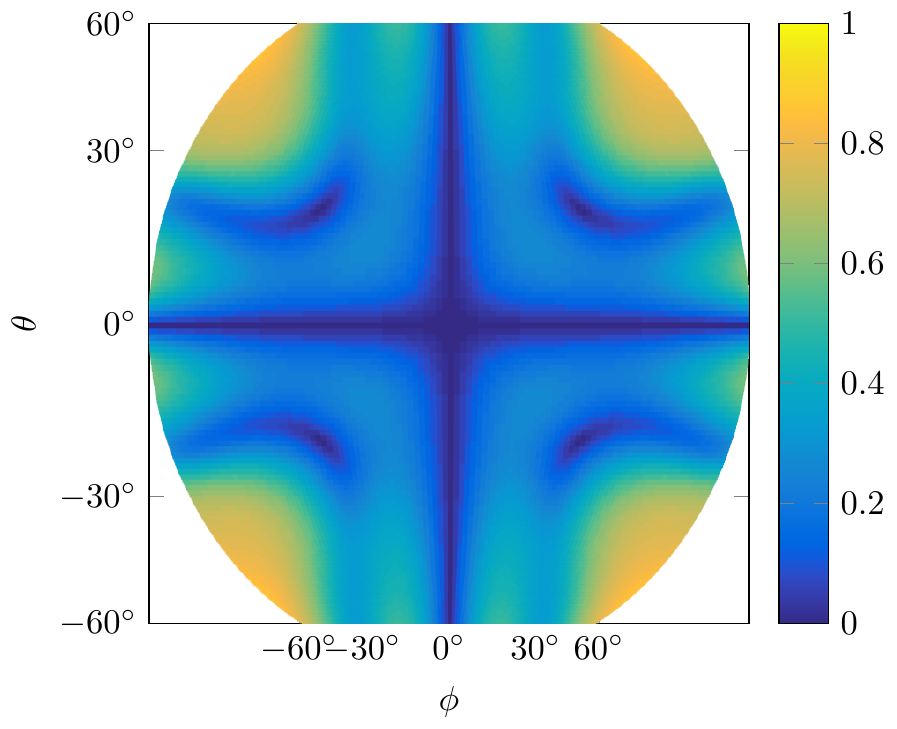}
	\end{center}\vspace{-4mm}
	\small\hspace{12mm}(e) $I_\text{a}$ at 2.8 GHz\hspace{22mm}(f) $I_\text{p}$ at 2.8 GHz\vspace{-1mm}
	\caption{\small Spatial distribution of polarization deficiencies $I_\text{a}$ and $I_\text{p}$ of the present 2-port antenna. Higher value means worse performance.}
	\label{fig:deficiency}
\end{figure}

The spatial distribution of $I_\text{a}$ and $I_\text{p}$ of the 2-port planar Eleven antenna is illustrated in Fig.~\ref{fig:deficiency} at different frequencies. Comparison of these plots with Fig.~\ref{fig:coverage}, clearly shows how the presence of the polarization deficiencies affect the spatial distribution of the MIMO coverage. In the presence of high polarization deficiency, more power is required at the transmitter side in order to acheive 95\% PoD, which means reduced efficiency.

\subsection{Dual-Polarized 4-port and 8-port MIMO Antenna}
As mentioned earlier, in addition to the 2-port mode, the planar Eleven antenna can be operated in 4-port and 8-port modes. In principle, more antenna ports can improve the MIMO efficiency by providing more diversity. Of course, this improvement is dependent on the correlation between the antenna ports, mutual coupling and embedded antenna efficiency.

\begin{figure}[!t]
	\begin{center}
		\includegraphics[width=0.8\columnwidth]{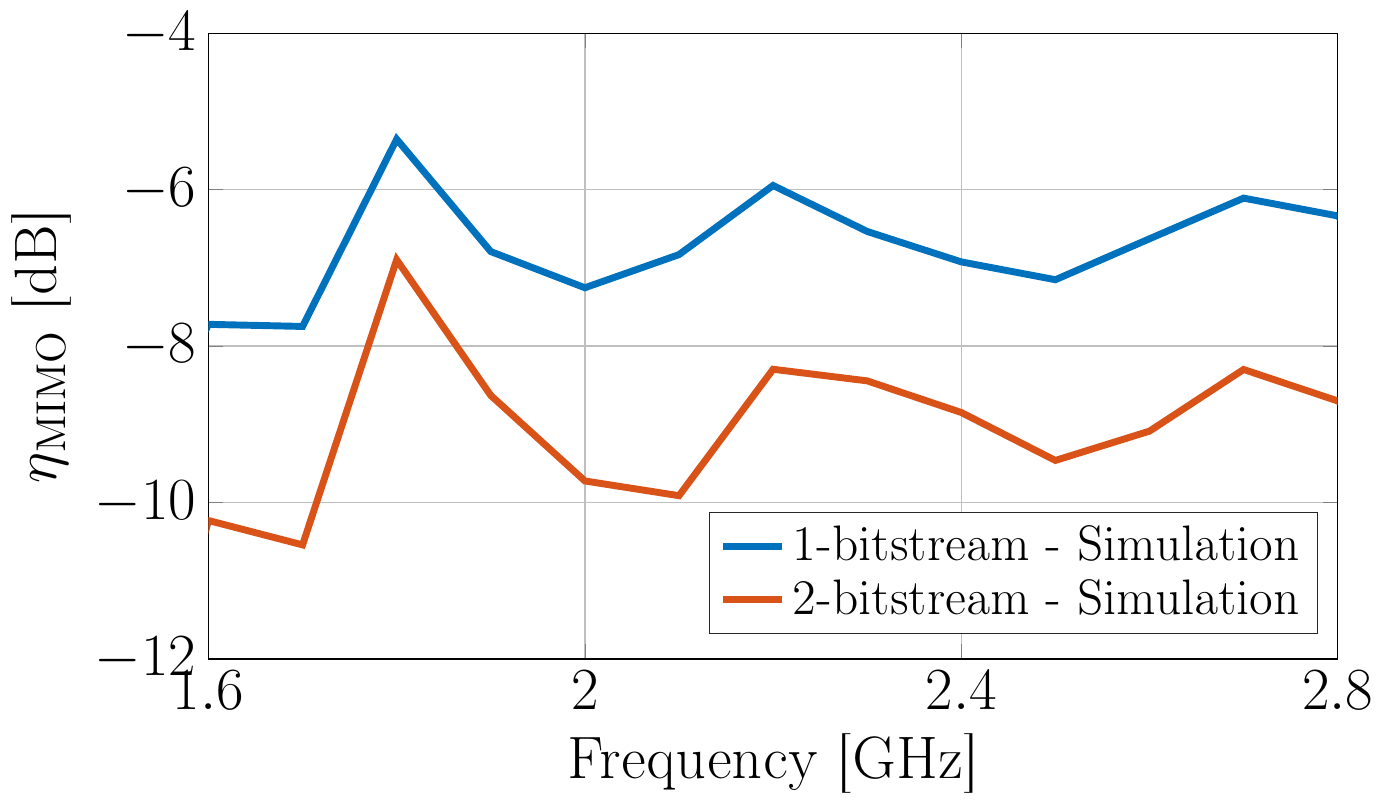}
	\end{center}
	\caption{\small 1-bitstream $4 \times 1$ and 2-bitstream $4 \times 2$ MIMO efficiency of the 4-port antenna in Random-LOS.}
	\label{fig:efficiency_4ports_rlos}
\end{figure}

\begin{figure}[!t]
	\begin{center}
		\includegraphics[width=0.8\columnwidth]{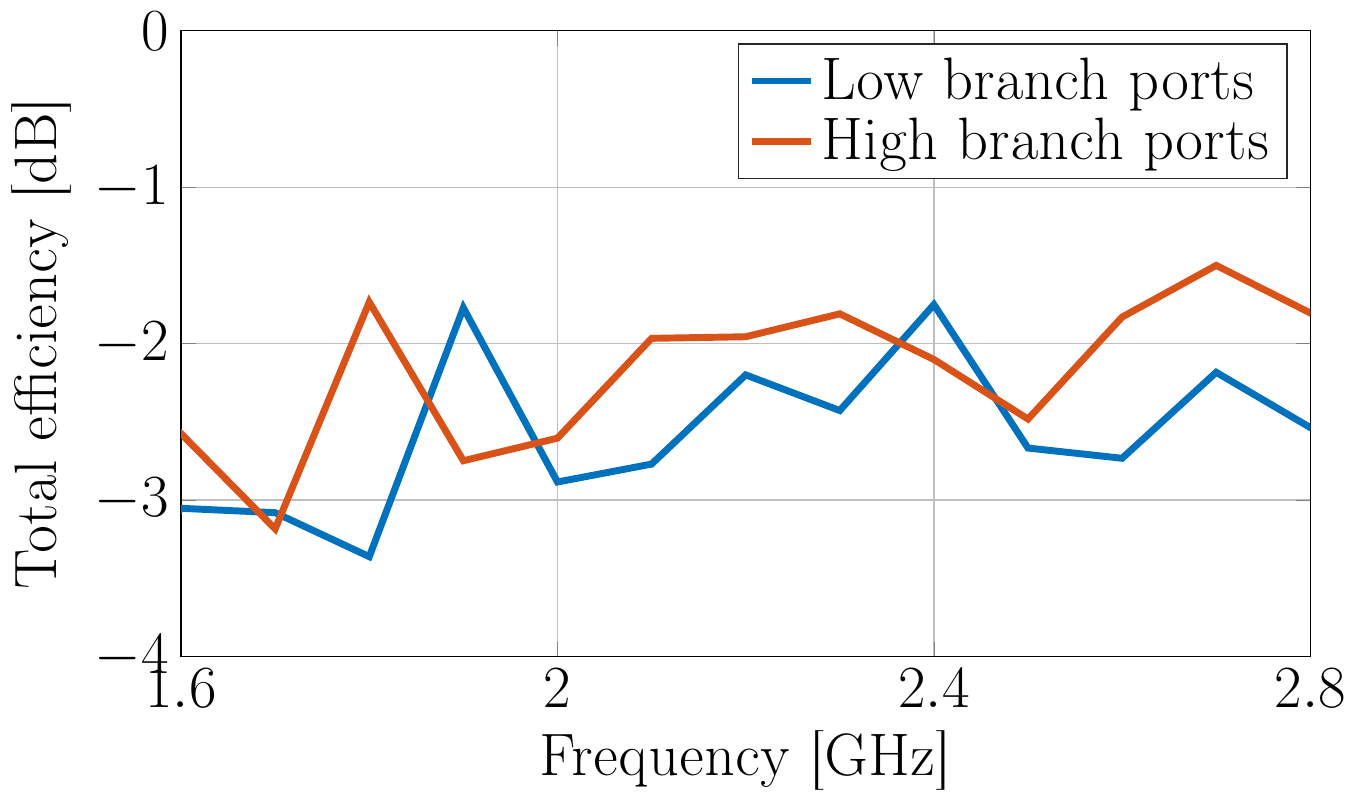}
	\end{center}
	\caption{\small Total embedded efficiency of the ports on low and high branches, 4-port antenna.}
	\label{fig:Total_efficiency_4ports}
\end{figure}

1- and 2-bitstream MIMO efficiency of the 4-port antenna in Random-LOS environment are plotted in Fig.~\ref{fig:efficiency_4ports_rlos}. Compared to Fig.~\ref{fig:efficiency_2ports_rlos}, it is evident that the MIMO efficiency improves by employing the antenna in 4-port mode of operation. Here, the total embedded antenna efficiency, including the reflection coefficient at the antenna ports is used to determine the MIMO efficiency. The total embedded efficiencies of ports on low and high branches, are plotted in Fig.~\ref{fig:Total_efficiency_4ports} for further reference.

\begin{figure}[!t]
	\begin{center}
		\includegraphics[width=0.8\columnwidth]{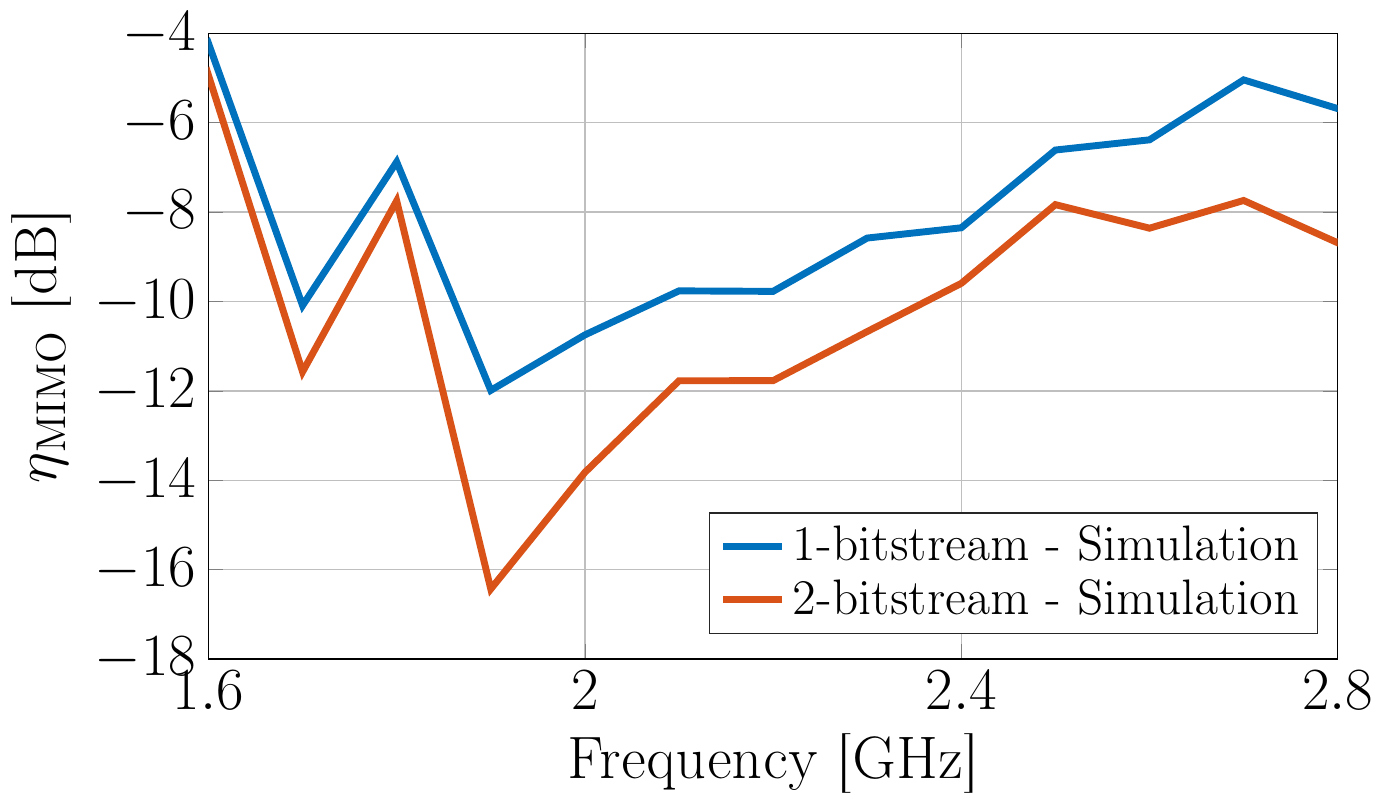}
	\end{center}
	\caption{\small 1-bitstream $8 \times 1$ and 2-bitstream $8 \times 2$ MIMO efficiency of the 8-port antenna in Random-LOS.}
	\label{fig:efficiency_8ports_rlos}
\end{figure}

\begin{figure}[!t]
	\begin{center}
		\includegraphics[width=0.8\columnwidth]{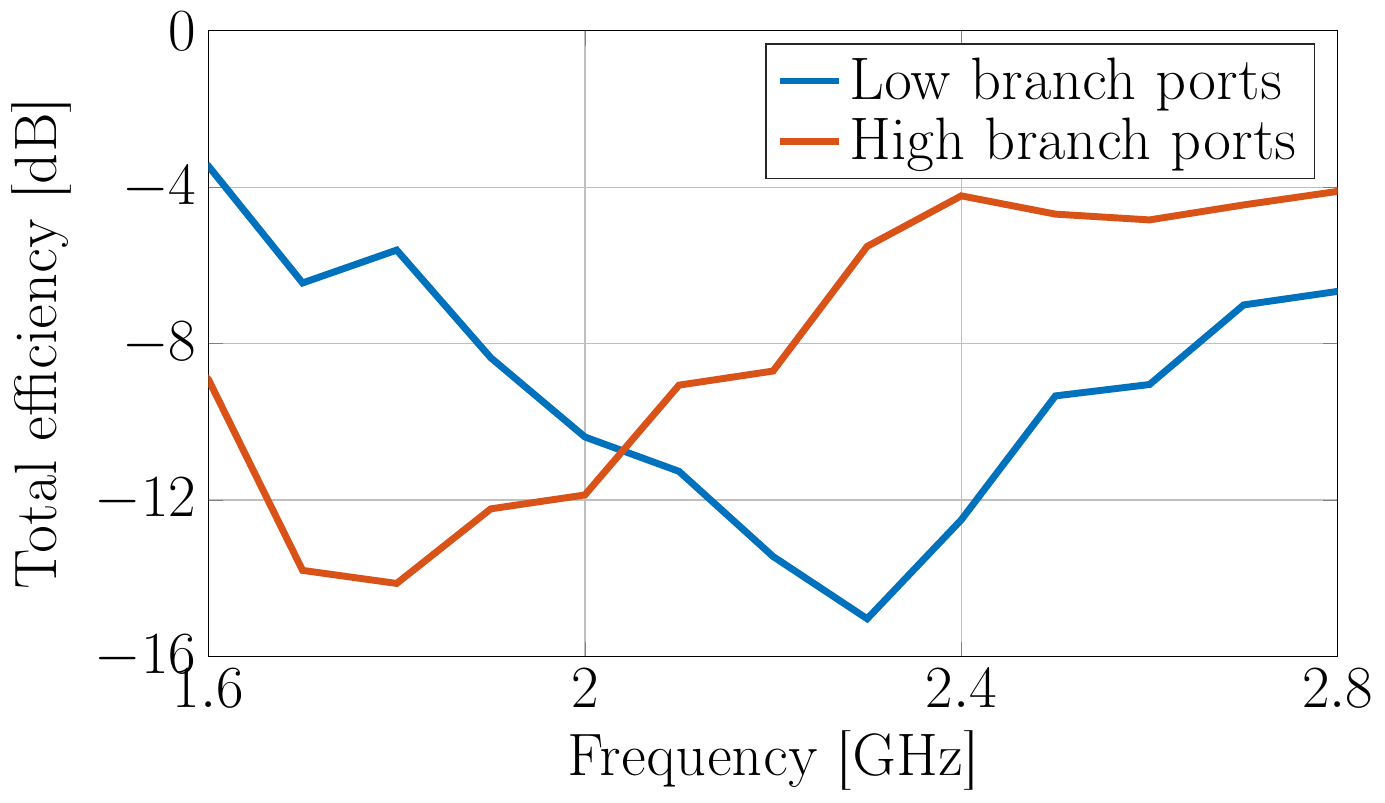}
	\end{center}
	\caption{\small Total embedded efficiency of the ports on low and high branches, 8-port antenna.}
	\label{fig:Total_efficiency_8ports}
\end{figure}

1- and 2-bitstream MIMO efficiency of the 8-port antenna in Random-LOS environment are plotted in Fig.~\ref{fig:efficiency_8ports_rlos}. Compared to Fig.~\ref{fig:efficiency_2ports_rlos}, it can be observed that the MIMO efficiency is generally degraded over the bandwidth of operation. This degradation is largely due to low total embedded efficiencies of the antenna ports. The total embedded efficiencies of ports on low and high branches in 8-port mode, are plotted in Fig.~\ref{fig:Total_efficiency_8ports} which clearly illustrates the effect of the antenna total embedded efficiency on the MIMO efficiency. The total embedded efficiency and MIMO efficiency of the 8-port antenna can be improved by addition of a proper matching circuit.

In comparison of the different operation modes of the antenna, we can conclude that the antenna performs best in the 4-port mode. Also in the 2-port mode, the performance of the antenna is acceptable. But using the present planar Eleven antenna in 8-port mode is not recommended.

\section{Conclusion}
A novel planar type Eleven Antenna is designed for micro base-station working in 1.6 GHz to 2.8 GHz frequency band. The flat structure of the antenna makes the manufacturing process simple and the low-profile makes it a suitable candidate for wall-mounted applications. The antenna can be operated in 2-, 4-, and 8-port modes, with small modifications. For micro base-stations, Random-LOS is more pronounced compared to RIMP, due to use of lower power and smaller cell size. The performance of the antenna is evaluated in Random-LOS and RIMP environments for an intended coverage of $120^\circ$ in both elevation and azimuth planes. The MIMO efficiency of the antenna has small variation over the frequency band. The spatial MIMO coverage of the antenna and the effect of polarization deficiencies are studied in the 2-port mode of operation. The 4-port mode provides higher MIMO efficiency due to increased diversity. However, the 8-port mode's MIMO efficiency is severely impaired due to sub-optimal matching and embedded antenna efficiency.

\ifCLASSOPTIONcaptionsoff
  \newpage
\fi

\bibliographystyle{IEEEtran}




\end{document}